\documentclass[11pt]{article}

\usepackage{graphicx}
\usepackage[squaren]{SIunits}
\usepackage{isolatin1}
\usepackage{hyperref}
\usepackage{xcolor}
\usepackage{caption}
\usepackage{cite}

\newcommand{\dd}{\mbox{d}}

\newcommand{\gfrac}[2]{\displaystyle\frac{#1}{#2}}

\newcommand{\erg}{\textrm{erg}}
\newcommand{\an}{\textrm{year}}
\newcommand{\few}{\textrm{few}}

\newcommand{\pitch}{\textrm{p}}
\newcommand{\mini}{\textrm{min}}
\newcommand{\maxi}{\textrm{max}}
\newcommand{\Aeff}{A_{\textrm{eff}}}
\newcommand{\paire}{\textrm{pair}}
\newcommand{\bisectrix}{\textrm{bisectrix}}
\newcommand{\photon}{\textrm{photon}}
\newcommand{\momentum}{\textrm{momentum}}

\definecolor{brun}{rgb}{0.39,0.26,0.18}
\newcommand{\Brun}[1]{\textcolor{brun}{#1}}

\newcommand{\Blue}[1]{\textcolor{blue}{#1}}
\newcommand{\Red}[1]{\textcolor{red}{#1}}

\newcommand{\Green}[1]{\textcolor{green}{#1}}
\newcommand{\Cyan}[1]{\textcolor{cyan}{#1}}
\newcommand{\Magenta}[1]{\textcolor{magenta}{#1}}

\begin{document} 

\title{On the Angular Resolution of Pair-Conversion $\gamma$-Ray Telescopes}

\author{D.~Bernard,
\\
LLR, Ecole Polytechnique, CNRS/IN2P3, 91128 Palaiseau, France}

\maketitle 

\begin{center}
 \large \textbf{Proceedings of a presentation to
 Session 14,
 ``Future spaceborne MeV detectors and related astrophysics'',
of the
 2023 International Conference of Deep Space Sciences,
April 2023, Hefei, China
}
\end{center}

\begin{abstract}
I present a study of the several contributions to the single-photon
angular resolution of pair telescopes in the MeV energy range.
I examine some test cases, the presently active {\sl Fermi} LAT, the
``pure-silicon'' projects ASTROGAM and AMEGO-X, and the emulsion-based
project GRAINE.
\end{abstract}

{\em keywords }:
$\gamma$-rays, pair production, telescope, angular resolution

\section{Introduction}

The {\sl Fermi} Large Area Telescope (LAT) \cite{Rando:2022zyj} has been
watching the $\gamma$-ray sky for 15 years, has provided invaluable
observations for over 6000 sources
(12 years'\cite{Fermi-LAT:2022byn}), and is monitoring daily the full
sky for transients possibly associated with gravitational wave
detections or very high energy (VHE) neutrino detections, over a wide
field of view
(with an $\approx 2.5 \, \meter^2 \steradian$ acceptance
for an $\approx 1 \, \meter^2$ effective area
above 1\,GeV \cite{Fermi:LAT:Performance}).
None of us is eternal, though, and it is time to prepare for the
$\gamma$-telescope of the next decade.

\begin{figure} [th]
 \begin{center}
\includegraphics[width=0.769\linewidth]{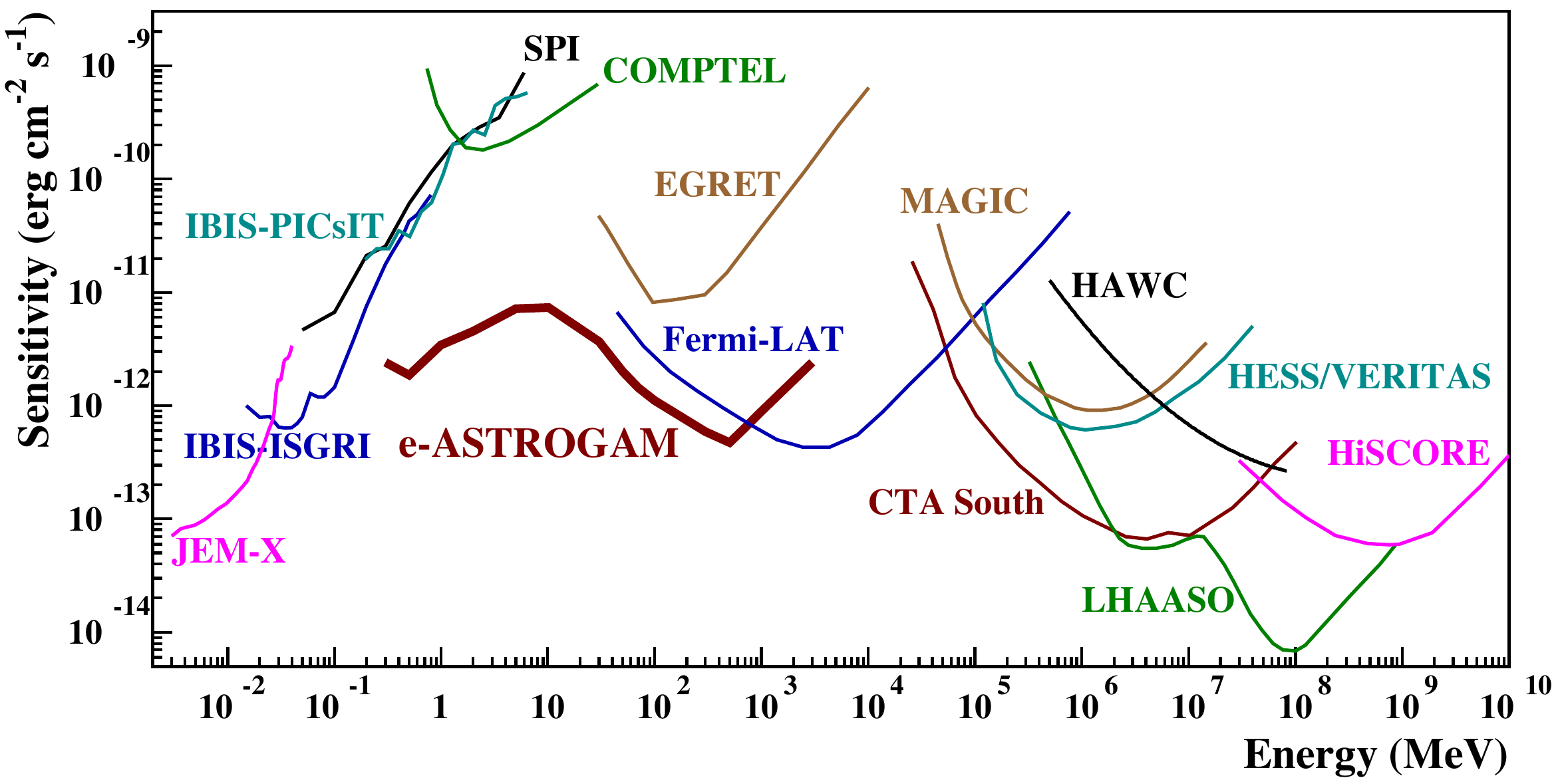}
\put(-320,40){\textbf{\Red{Compton}}}
\put(-220,40){\textbf{\Red{Pairs}}}
\put(-272,133){\textbf{\Red{Gap}}}
\caption{\label{fig:Diff:Sens} \sl 
 Expected differential continuum sensitivity
 of the ASTROGAM project (\Brun{\bf brown curve}), compared to that 
 of past and present X-ray and $\gamma$-ray telescopes.
 The Compton and Pair parts of the $\gamma$-ray realm are indicated in
 \textbf{\Red{red}}, so is the \textbf{\Red{sensitivity gap}} between them.
 Adapted from \cite{e-ASTROGAM:2017pxr}. }
 \end{center}
\end{figure}

\begin{figure} [th]
 \includegraphics[width=0.505\linewidth]{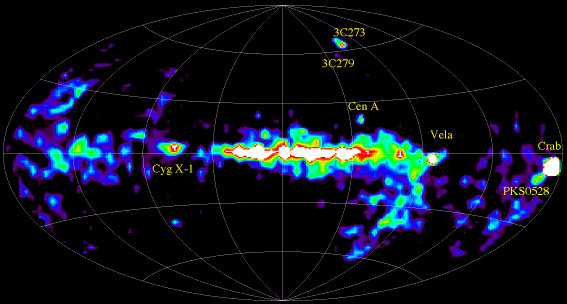}
 \hfill
 \includegraphics[width=0.485\linewidth, trim=4 100 4 180, clip]{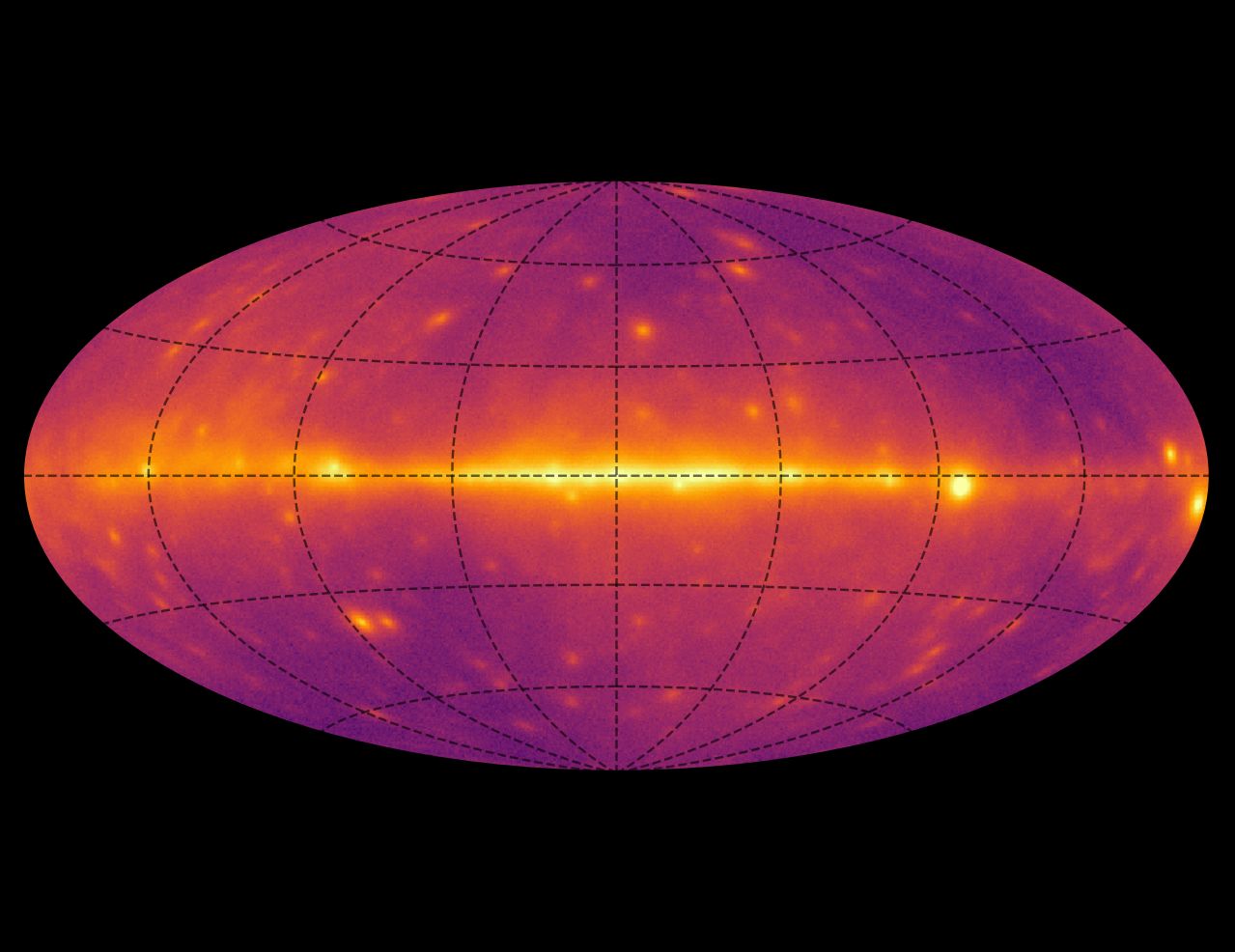}
\caption{\label{fig:Sky:Maps} \sl 
 Maps of the $\gamma$-ray sky from the COMPTEL
 (left, Compton regime, 1 -- 30\,MeV, 9 years, \cite{COMPTEL})
 and from the LAT
 (right, pair creation regime, 20 -- 200\,MeV, 13 years, \cite{LAT-20-200})
 telescopes. }
\end{figure}

At lower energies, X-ray instruments such as the spectrometer SPI and
the imager IBIS on Board the INTEGRAL Satellite \cite{integral},
provide measurements with a similar sensitivity
(Fig. \ref{fig:Diff:Sens}).
The situation is much less favorable, though, in the energy range
0.2\,MeV -- 0.4\,GeV, that is, over three orders of magnitude, for
which no instrument is available with a sensitivity lower (better) than
$10^{-12}\erg \,\centi\meter^{-2}\second^{-1}$, 
while the last-century COMPTEL peaked above 
$10^{-10}\erg \,\centi\meter^{-2}\second^{-1}$.
A consequence of the terrible limitation of the sensitivity of the
available instruments is the terribly limited number of sources that
can be studied in the MeV energy range (Fig. \ref{fig:Sky:Maps}).
A number of instruments are being developed though, so as to study the
$\gamma$-ray sky in the future post-LAT era, and to fill the MeV
sensitivity gap.

\begin{figure} [th]
\begin{center}
 \includegraphics[trim=4 4 4 4 ,clip,width=0.645\textwidth]{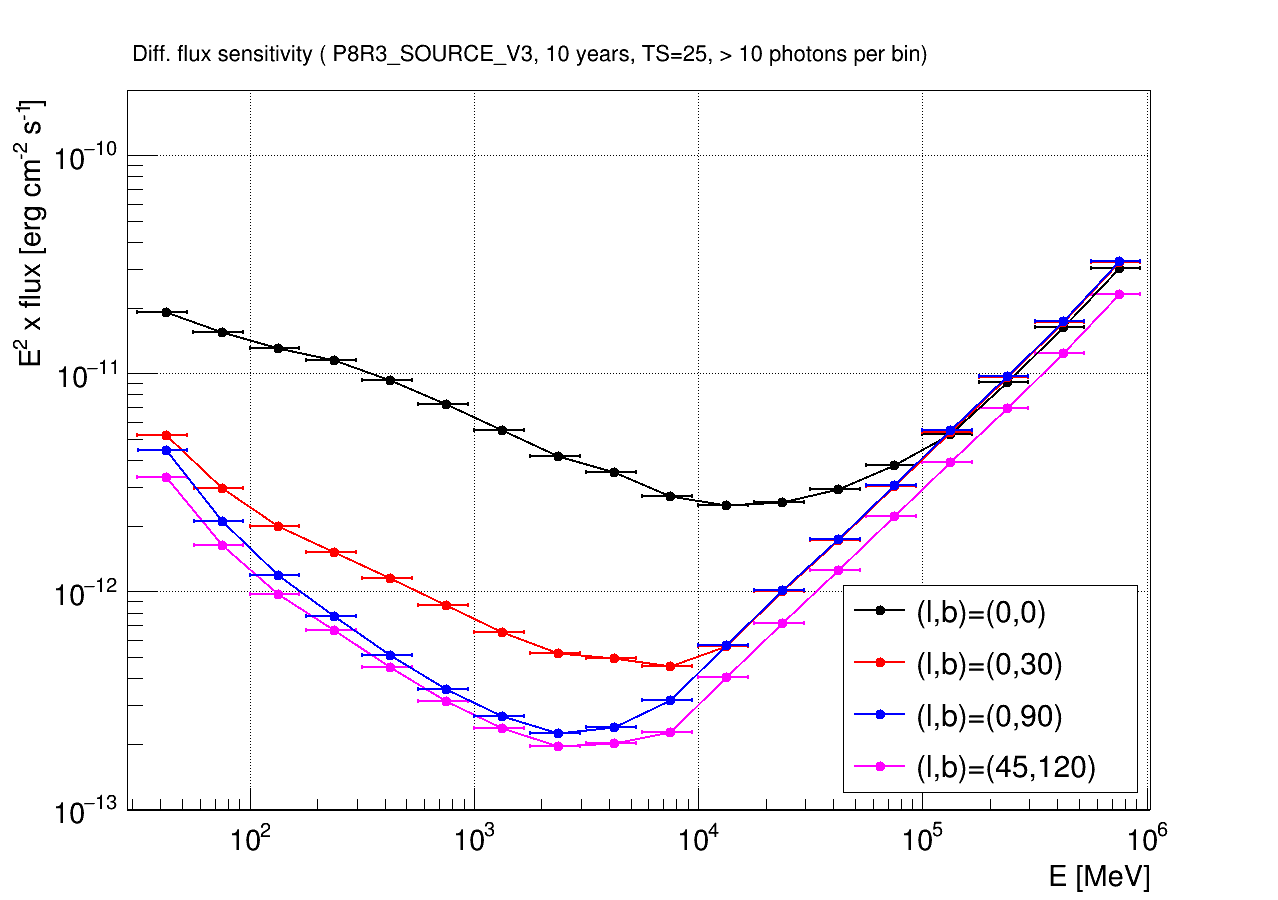}
\thicklines
\put(-10,100){\vector(-3,1){60}}
\put(-7,100){\scriptsize $N_\gamma > 10$}
\put(-28,119){\scriptsize Limited by photon statistics}
\put(-330,154){\vector(3,-1){70}}
\put(-455,154){\scriptsize Limited by background rejection}
\put(-455,112){{\scriptsize 
$s \approx 
\displaystyle \frac{n ~ \Red{\sigma_{\theta}} ~ E^2} {\Delta E} 
\sqrt{\frac{\pi\displaystyle \int B(E) ~ \dd E}{T ~ \Aeff}}$
}}
\put(-455,84){{\scriptsize $\Aeff$, effective area}}
\put(-455,63){{\scriptsize $T$, duration}} 
\put(-7,56){\scriptsize $b$ galactic latitude}
\end{center}
\caption{\label{fig:LAT:sens} \sl 
 Differential sensitivity of the {\sl Fermi} LAT
 \cite{Fermi:LAT:Performance}.}
\end{figure}

The so-called differential flux sensitivity, $s$, is the minimum flux
needed to get a $n$-standard-deviation detection from a point-like
gamma-ray source, estimated for a data taking of duration $T$ ($n=3$,
$T\,=\,1\,\an$ for the project ASTROGAM, Fig. \ref{fig:Diff:Sens} and
\cite{e-ASTROGAM:2017pxr}).
%
%
Figure \ref{fig:LAT:sens} shows the variation of $s$ with $E$ for the
{\sl Fermi} LAT ($n=5$, $T\,=\,10\,\an$) \cite{Fermi:LAT:Performance}.
Besides an increase of the effective area, $\Aeff$, and/or of the
mission duration, $T$, both of which depend heavily on the funding
availability, the only way to improve on the sensitivity (decrease the
value of $s$) is to improve on the angular resolution,
$\Red{\sigma_{\theta}}$, something that has been the topic of my talk.

``We are not facing a sensitivity wall, we are facing an angular
resolution wall'' (I. Grenier, 2016).

Upon the conversion of an incident $\gamma$-ray ``in the field'' of a
charged target, supposed to be at rest, a pair of $e^+$ and $e^-$
leptons is created, with momenta
$\vec p_+$ and 
$\vec p_-$, and some momentum, $\vec q$ is transferred to the target that then ``recoils'':
\begin{equation}
 \vec k = \vec p_+ + \vec p_- + \vec q.
\end{equation}

The astronomer reconstructs the photon momentum $\vec k$ from the
measurement of the momenta of the particles in the final state.
 \begin{itemize}
\item 
When the target is the nucleus of an atom of the detector
(``nuclear conversion''),
the recoil track length is too short that $\vec q$ can be measured,
even with a low-density (gas) tracker.
\item 
When the target is an electron
(``triplet conversion''),
the track can be longer, but the cross
section is lower and in practice the fraction of the events that has a
recoil momentum large enough to enable a measurement is tiny
(see, eg., Fig. 6 of \cite{Bernard:2013jea}).
 \end{itemize}
So (\cite{Bernard:2012uf} and references therein), the single photon angular resolution can be broken down into the following contributions:
 \begin{itemize}
\item (1) the fact that $\vec q$ cannot be measured, a missing
 piece in the calculation of $\vec k$;
\item (2) 
 the momentum resolution for each track of the pair;
\item (3)
 the single-track angular resolution.
 \end{itemize}

\begin{figure} [th]
\begin{center}
\includegraphics[width=0.9\linewidth]{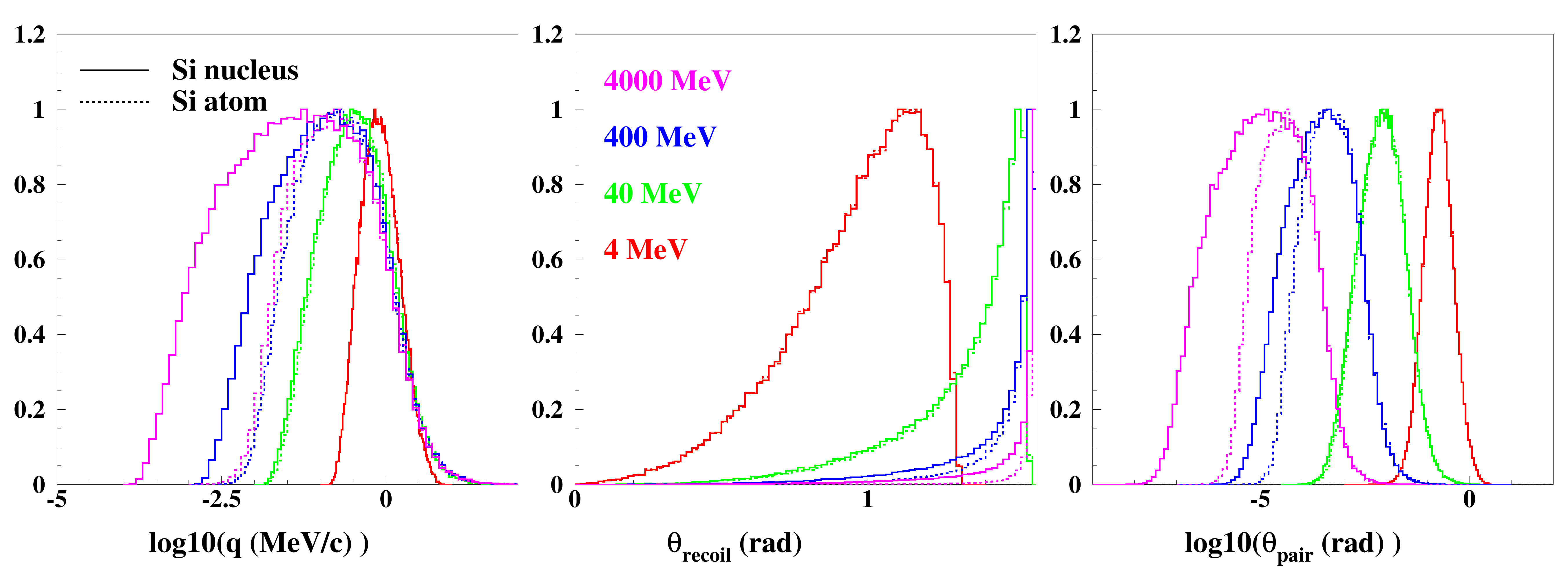}
\put(-160,9){\small ${\gfrac{\pi}{2}}$}
\end{center}
\caption{\label{fig:recoil} \sl 
 \sl Target recoil
 in nuclear conversions of $\gamma$-rays on silicon, either on
 isolated nuclei (solid curves) or on full atoms (dotted curves), for
 several incident photon energies.
Left, (log of) recoil momentum magnitude.
Center, recoil polar angle.
Right, (log of) the (true, not the $q/k$ approximation) contribution
of the missing recoil to the single-photon angular resolution. }
\end{figure}

\section{Recoil}

The direction of the recoil is found to be, asymptotically at high
energy, transverse to the direction of the incident photon
(see center plot in Fig. \ref{fig:recoil}), so the induced angular
shift of the reconstructed photon due to the missing recoil is of
about\footnote{The contribution to the angular resolution,
in the $\theta_{68} \approx q_{68} /k$ approximation, is found to be
compatible with the exact value above a couple of MeV (Fig. 3 (right) of
\cite{Gros:2016zst}).}
$q / k$.
The distribution of the magnitude of the recoil momentum extends, in
principle, asymptotically, from $q_{\mini} = 2 m^2 / k$ up
to\footnote{
Under the approximation of $k \ll M c^2$, $M$ nucleus mass, that is
legitimate in the context of the ``MeV'' session of this conference.}
$q_{\maxi} = 2 k $.
In practice, though,
\begin{itemize}
 \item
the fraction of the cross section that extends above 1\,MeV/$c$ is
extremely small
 (nuclear conversion, see left plot of Fig. \ref{fig:recoil};
 triplet conversion, Fig. 6 of \cite{Bernard:2013jea});
 \item
 at high energies ($E > \few \times 100 \,\mega\electronvolt$), the lowest part
 of the $q$ spectrum is suppressed by the screening of the electric
 field of the target by the electron ``cloud''
 (dotted curves on the left plot of Fig. \ref{fig:recoil}).
\end{itemize}

\begin{figure} [th]
\begin{center}
\includegraphics[width=0.75\linewidth]{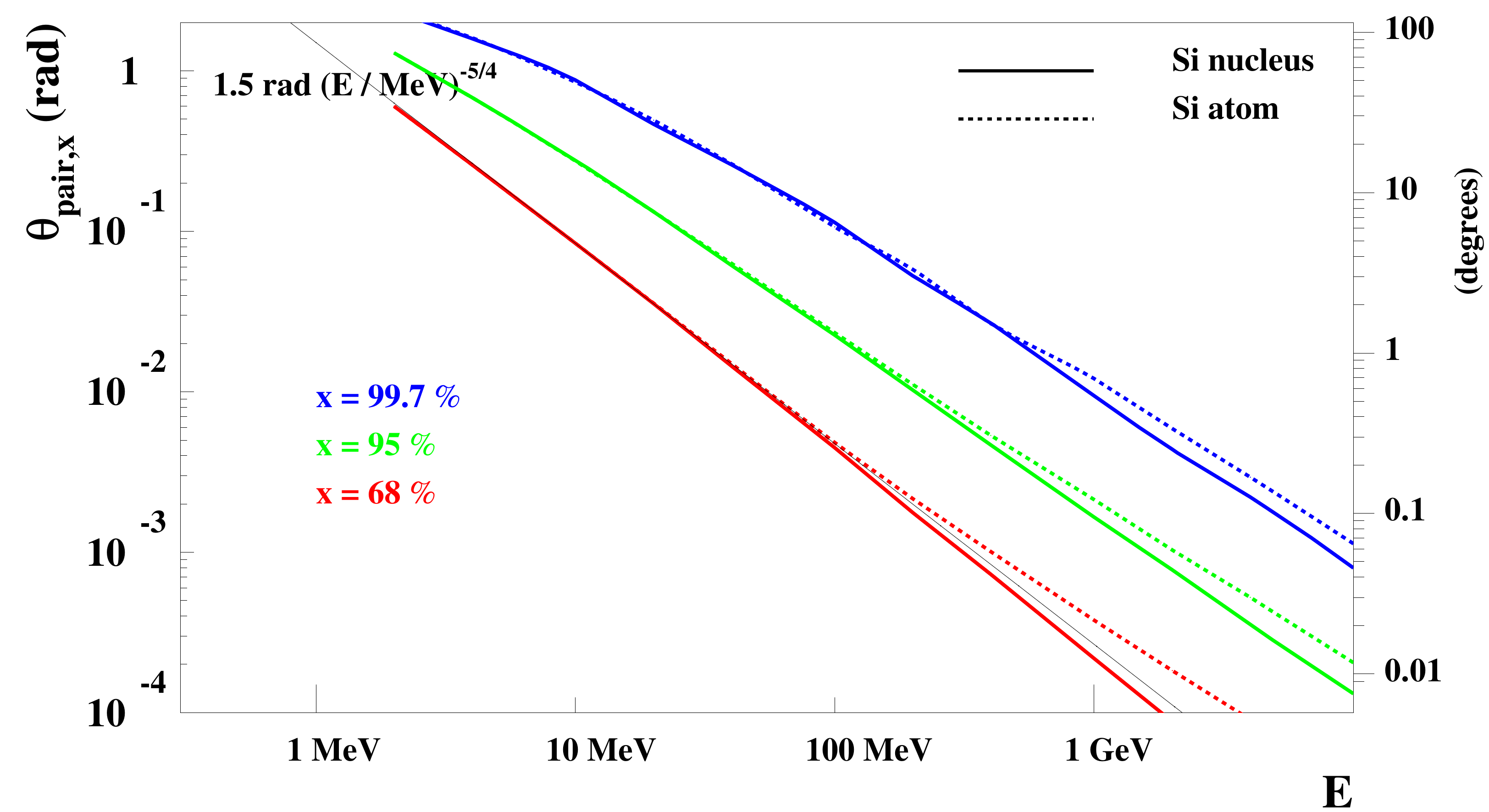}
\end{center}
\caption{\label{fig:recoil:containment} \sl 
Variation of the contribution of the missing recoil momentum to the
single-photon angular resolution for $\gamma$-ray conversions to pairs
on silicon (solid line, isolated nucleus; dotted line, full atom).
Values at $x = \Red{68}, \Green{95} ~ \textrm{and} ~ \Blue{99.7}\,\%$
``containment'' are shown.
The thin black line is a good representation of the variation for $x = \Red{68}\,\%$
\cite{Bernard:2012uf,Gros:2016zst}.
}
\end{figure}

As the angular kick induced by the missing recoil in the
reconstruction of the photon direction shows a strongly non-Gaussian
distribution, the contribution to the angular resolution is refered to
as the angle value that ``contains'' a fraction $x$ of the events,
with $x = \Red{68}, \Green{95} ~ \textrm{and} ~ \Blue{99.7}\,\%$
(Fig. \ref{fig:recoil:containment} and 
\cite{Bernard:2012uf,Gros:2016zst}).
The $\Red{68}\,\%$ containment value can be parametrized by
$\theta_{68} = 1.5 \, \radian (E / \mega\electronvolt)^{-5/4}$
(the thin black line on Fig. \ref{fig:recoil:containment})
\cite{Bernard:2012uf,Gros:2016zst}, and amounts to 
$ 0.27^\circ$ at $100 \,\mega\electronvolt$.

A number of misconceptions have been published in the past on this
kinematical limit of the angular resolution of pair telescopes.
Recently, for example, \cite{Aboudan:2022ver} states
(eq. (30), up to a trigonometric function) a $1/E$ variation, based on
the allegation that
``The most probable energy transfer to the nucleus is close to the
energy at rest of the electron''.
Actually \cite{Aboudan:2022ver} uses the Geant4
{\sl G4EmLivermorePolarizedPhysics} list, in which, in the
{\sl G4LivermorePolarizedGammaConversionModel} physics model,
the Bethe-Heitler differential cross section\cite{Bethe-Heitler,BerlinMadansky1950,May1951} is not sampled;
instead the polar angles of the two leptons are taken at random
independently, so given the $1/E$ (asymptotic) scaling of the
distribution of the opening angle \cite{Olsen1963}
(see also Fig. \ref{fig:open}, left), the induced violation of the energy-momentum
conservation in the conversion ends up in an artificial,
pseudo-kinematical contribution to the angular resolution that goes
like $1/E$ (Figs. 10, 12 and 14 of \cite{Aboudan:2022ver}).

Similar comments can be made on alternative physics models such as
{\sl G4BetheHeitlerModel}
 or \newline
{\sl G4LivermoreGammaConversionModel} 
for which energy (not momentum) conservation is used to correlate the
polar angles of the two leptons in a coplanar, recoil-less, scheme.
In that case, the induced $q_{68}$ obviously collapses to
zero (asymptotically),
(magenta \Magenta{open circles} in Figs. 6-8 of \cite{Gros:2016zst}).

I would like to respectfully recommend that that tools, in general,
and event generators, in particular, be validated before they are used,
given the specifications of the intended
use\footnote{As analytical expressions of the distributions of the
 single track polar angle and of the fraction of the energy carried
 away by (e.g.) the positron are available
\cite{Bethe-Heitler}, the physics models that I
 have commented on actually provide excellent simulations of the
 properties of electromagnetic showers.}.
See, eg., how the method used used in {\sl G4BetheHeitler5DModel} to
sample the full (5 dimensional) Bethe-Heitler differential cross
section \cite{Bethe-Heitler,BerlinMadansky1950,May1951}
was validated \cite{Bernard:2018hwf,Semeniouk:2019cwl}
before it was chosen \cite{Geant4:2019cxv}
by Geant4 \cite{Agostinelli:2002hh,Allison:2016lfl} as the default
physics model for the simulation of precision electromagnetic physics.
For the recoil momentum distribution, which is the key point in the
present discussion, the simulated distribution was validated with
respect to the high-energy asymptotic expression from Jost {\sl et al.} 
\cite{Jost:1950zz} \footnote{With Borsellino's correction \cite{Borsellino1953} applied.}, see
Fig. 4 of \cite{Bernard:2012uf} and
Fig. 2 left of \cite{Geant4:2019cxv}.

\section{Single-Track Momentum Resolution}

Let's have a look at the final state in the plane perpendicular to the
direction of the incident photon

\hfill
\includegraphics[width=0.22\linewidth,trim=4 40 4 40 ,clip]{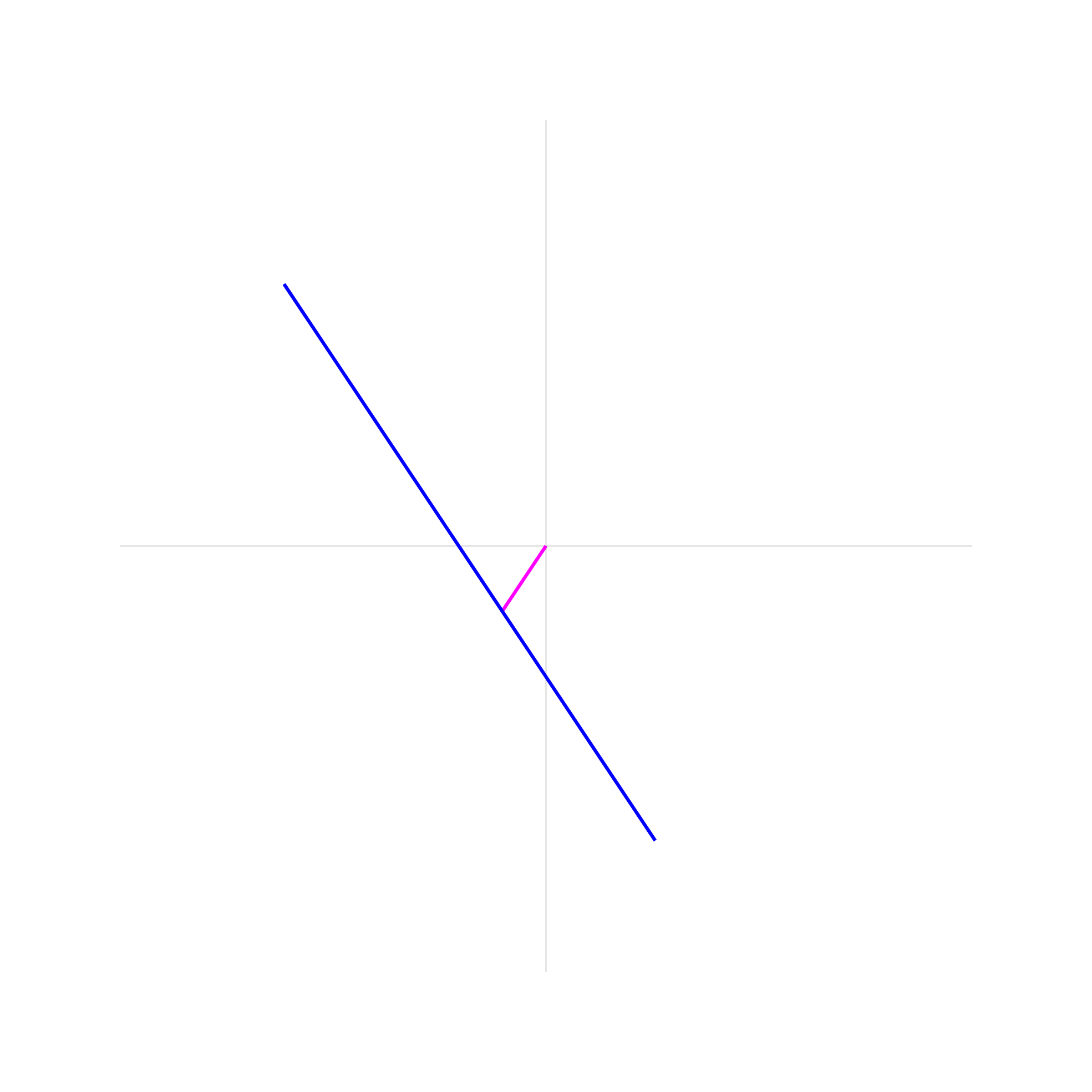}
\put(-44,11){\Blue{$e^-$}}
\put(-91,80){\Blue{$e^+$}}
\put(-80,35){\small \Magenta{pair}}
\put(-55,53){\footnotesize $\mathbf{\gamma}$}
\put(-7,49){\footnotesize{$\theta_x$}}
\put(-55,95){\footnotesize{$\theta_y$}}
\put(-87,49){\rotatebox[origin=c]{32}{ \resizebox{1.7cm}{!}{ \Blue{ \Bigg\} }}}}
\put(-30,65){\Blue{\small $\theta_{+-}$}}
\hfill
~

\begin{figure} [th]
\begin{center}
\includegraphics[width=0.36\linewidth]{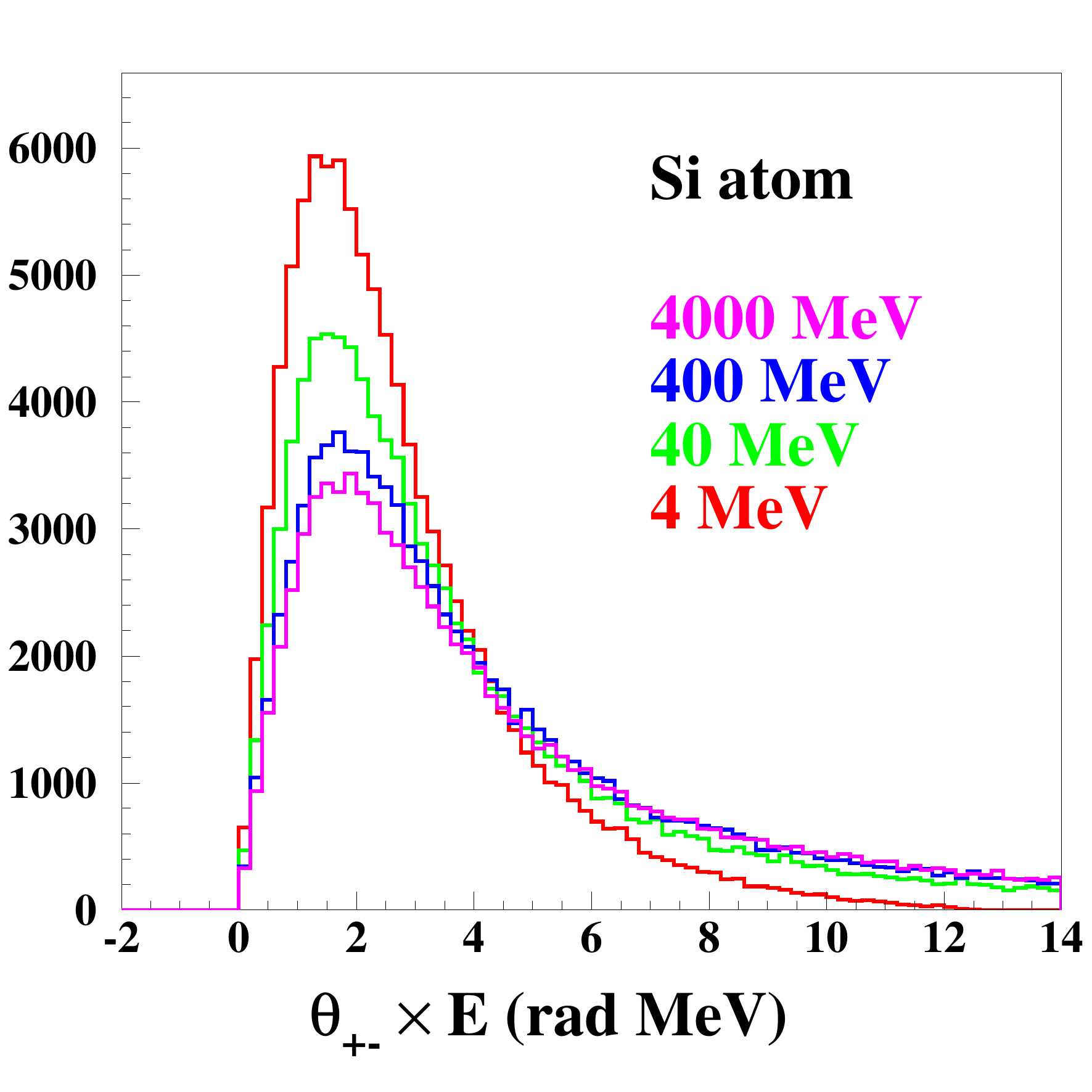}
\includegraphics[width=0.62\linewidth]{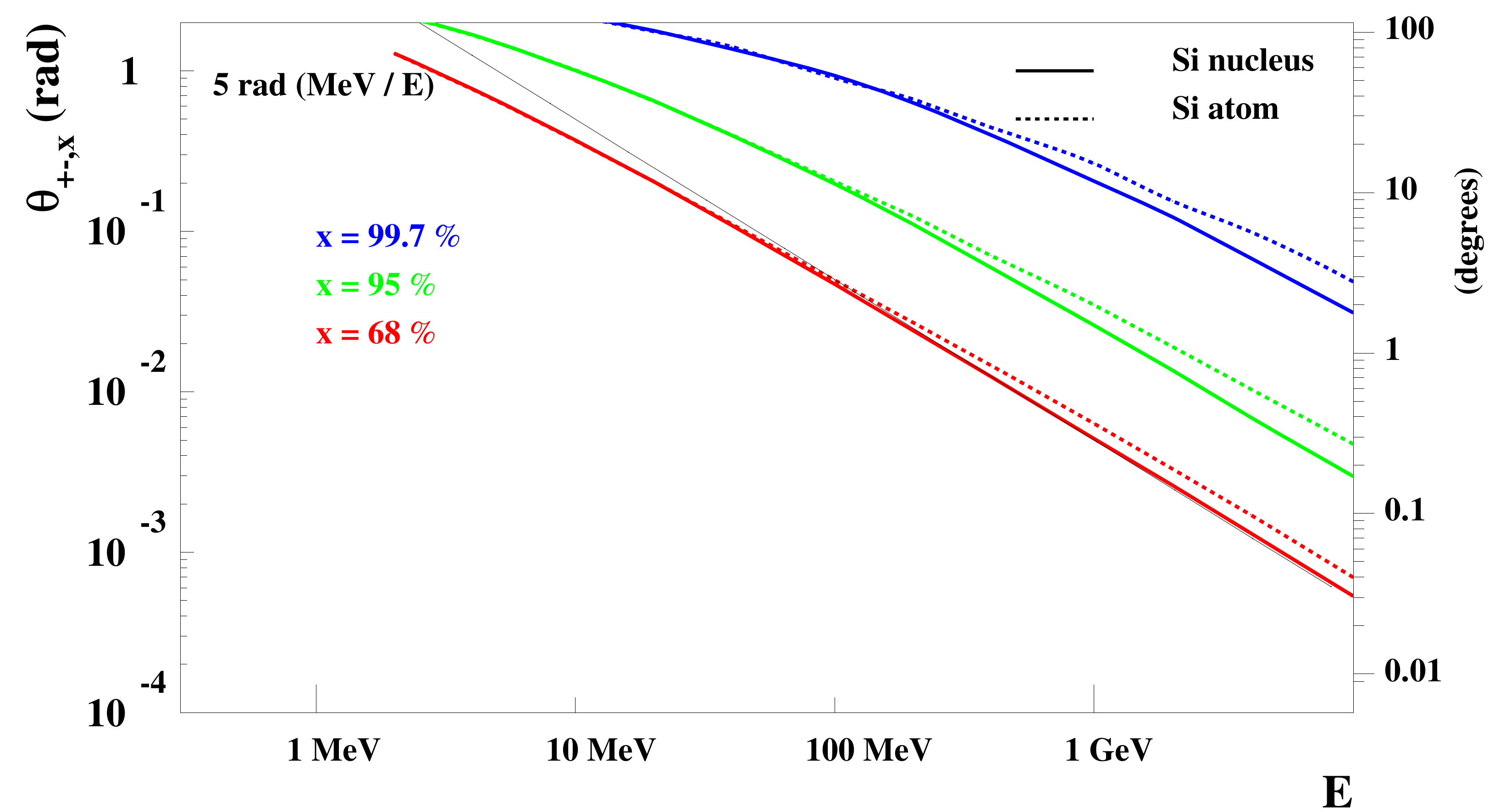}
\end{center}
\caption{\label{fig:open} \sl 
Pair opening angle, $\theta_{+-}$, in $\gamma$-ray conversions on silicon.
 Left, distribution of $\theta_{+-} \times E$, for several values of $E$.
 Right, $\theta_{+-}$ ``containment'' value as a function of $E$.
The thin black line is a good representation of the variation for $x = \Red{68}\,\%$.
 }
\end{figure}

The thin magenta \Magenta{segment} shows the transverse kick given to
the pair, back-to-back to the recoil.
Unless the magnitude of the momenta of the two tracks are measured, the 
direction of the reconstructed photon can lie anywhere on the thin blue
\Blue{segment} that extends from the \Blue{$e^+$} to the \Blue{$e^-$}, and
the extension of which is equal to the opening angle,
\Blue{$\theta_{+-}$}.
The distribution of $\theta_{+-}$ scales like $1/E$ asymptotically
(\cite{Olsen1963} and Fig. \ref{fig:open} left), as we have already
seen, and unfortunately it's magnitude is huge,
$5 \, \radian (\mega\electronvolt / E)$ at 68\,\% containment
(Fig. \ref{fig:open} right).

\subsection{Equipartition, bisectrix }

\begin{figure} [th]
\begin{center}
\includegraphics[width=0.82\linewidth]{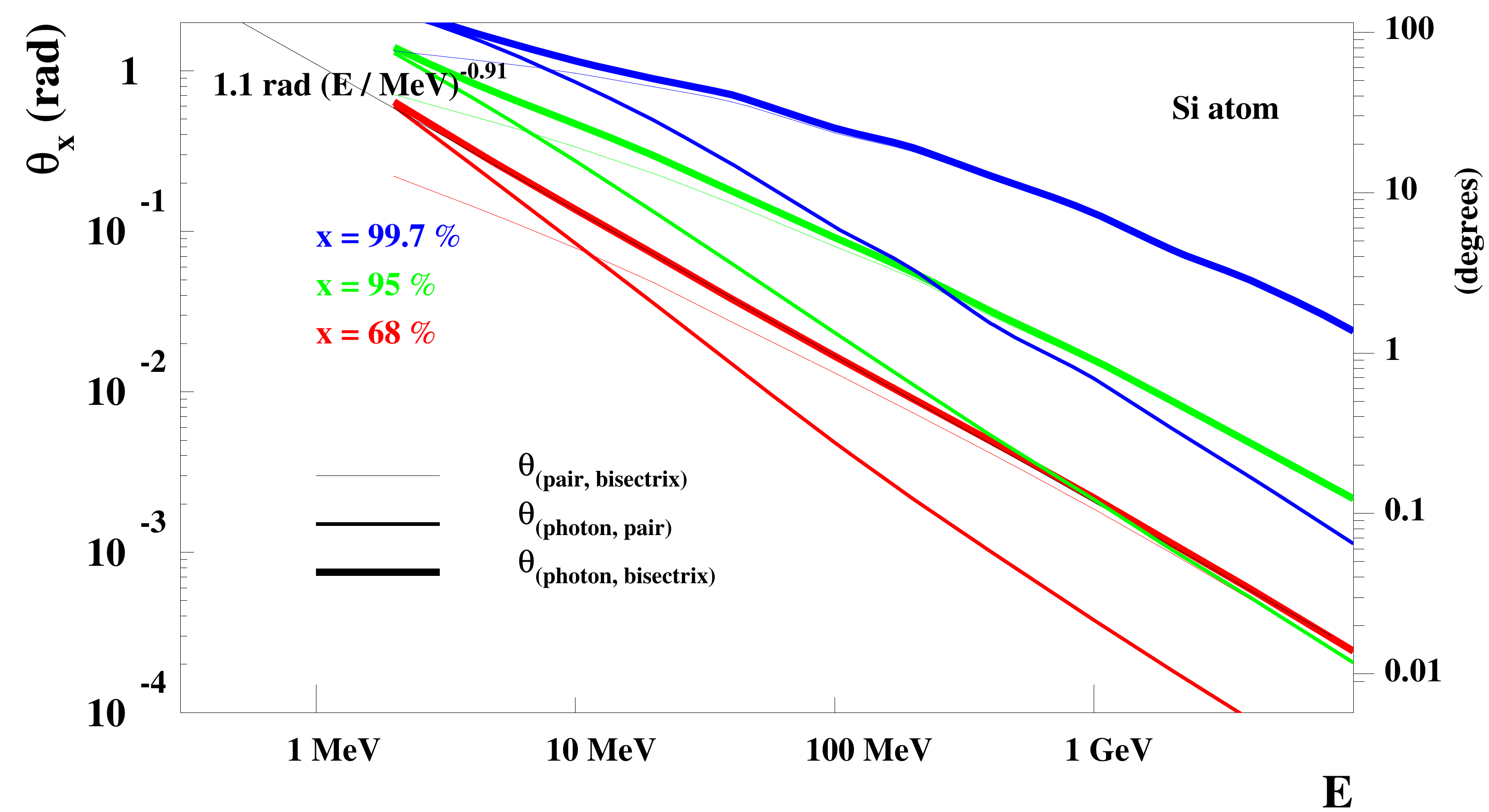}
\end{center}
\captionsetup{singlelinecheck=off}
\caption[hop]{\label{fig:bisectrix} \small \sl
Contributions to the single-photon angular resolution assuming
equipartition ($p_+ = p_-$).
\begin{itemize}
\item 
The angle between the true and the reconstructed pair momentum
 ($\theta_{(\paire,\bisectrix)}$, thin curve),
\item 
the angle between the true pair momentum and the incident photon direction
($\theta_{(\photon,\paire)}$, medium thickness),
\item 
the angle between the reconstructed momentum and the incident photon direction
($\theta_{(\photon,\bisectrix)}$, thick curve).
\end{itemize}
Values at $x = \Red{68}, \Green{95} ~ \textrm{and} ~ \Blue{99.7}\,\%$ ``containment'' are shown.
The thin black line is a good representation of the variation for $x = \Red{68}\,\%$.
}
\end{figure}

The easiest way to deal with the issue is to assume equipartition and
therefore to assign the bisectrix of the directions of the two leptons
as the direction of the reconstructed photon.
Figure \ref{fig:bisectrix} shows the contributions to the angular
resolution (containment values) of 
\begin{itemize}
\item
 the angle between the true and the reconstructed pair momentum
 assuming equipartition
 ($\theta_{(\paire,\bisectrix)}$, thin curve), 
\item the already mentioned contribution of the missing recoil momentum,
 ie., 
the angle between the true pair momentum and the incident photon direction
 ($\theta_{(\photon,\paire)}$, medium thickness)

 and their combination, into
\item the full angular resolution, ie.,
 the angle between the reconstructed pair momentum assuming equipartition and the incident photon direction
 ($\theta_{(\photon,\bisectrix)}$, thick curve).
\end{itemize}

We see that $\theta_{(\paire,\bisectrix)}$ dominates over
$\theta_{(\photon,\paire)}$, for $E > 10 \,\mega\electronvolt$.

\subsection{Measurement of the magnitude of the track momentum}

Several methods can be used, though, external to the
tracker/converter, to measure either the momentum or the energy of
each track\footnote{I assume that the photon energy resolutions
 quoted in the following table
 apply also to each track independently.},
but at the cost of a heavy impact of the mass budget of the mission:

\small
\hspace{-1cm}
\begin{tabular}{llllll}
 & \textbf{Calorimeter} & \textbf{Calorimeter} & \textbf{Calorimeter} & \textbf{Magnetic spectrometer} & 
 \\
 & {\sl Fermi} LAT \cite{Fermi:LAT:Performance} & ASTROGAM \cite{e-ASTROGAM:2016bph} & AMEGO \cite{Caputo:2022xpx} & AMS-02 \cite{Beischer:2020rts}
\\
 & CsI(Tl) ($8.6 \, X_0$) & CsI(Tl) ($4.3 \, X_0$) & CsI(Tl) & Magnet + Si tracker
\\
 & $\sigma_E / E = 20\,\%$ @ 100\,MeV
 & $\sigma_E / E = 20 - 30\,\%$ @ $>$30\,MeV
 & 17\% at 100 MeV
 & $\sigma_p / p = 14\,\%$ @ 200\,MeV
\\
& (68\,\%) & & (68\% containment half width)
\end{tabular}
\normalsize

\begin{figure} [th]
\begin{center}
\includegraphics[width=0.82\linewidth]{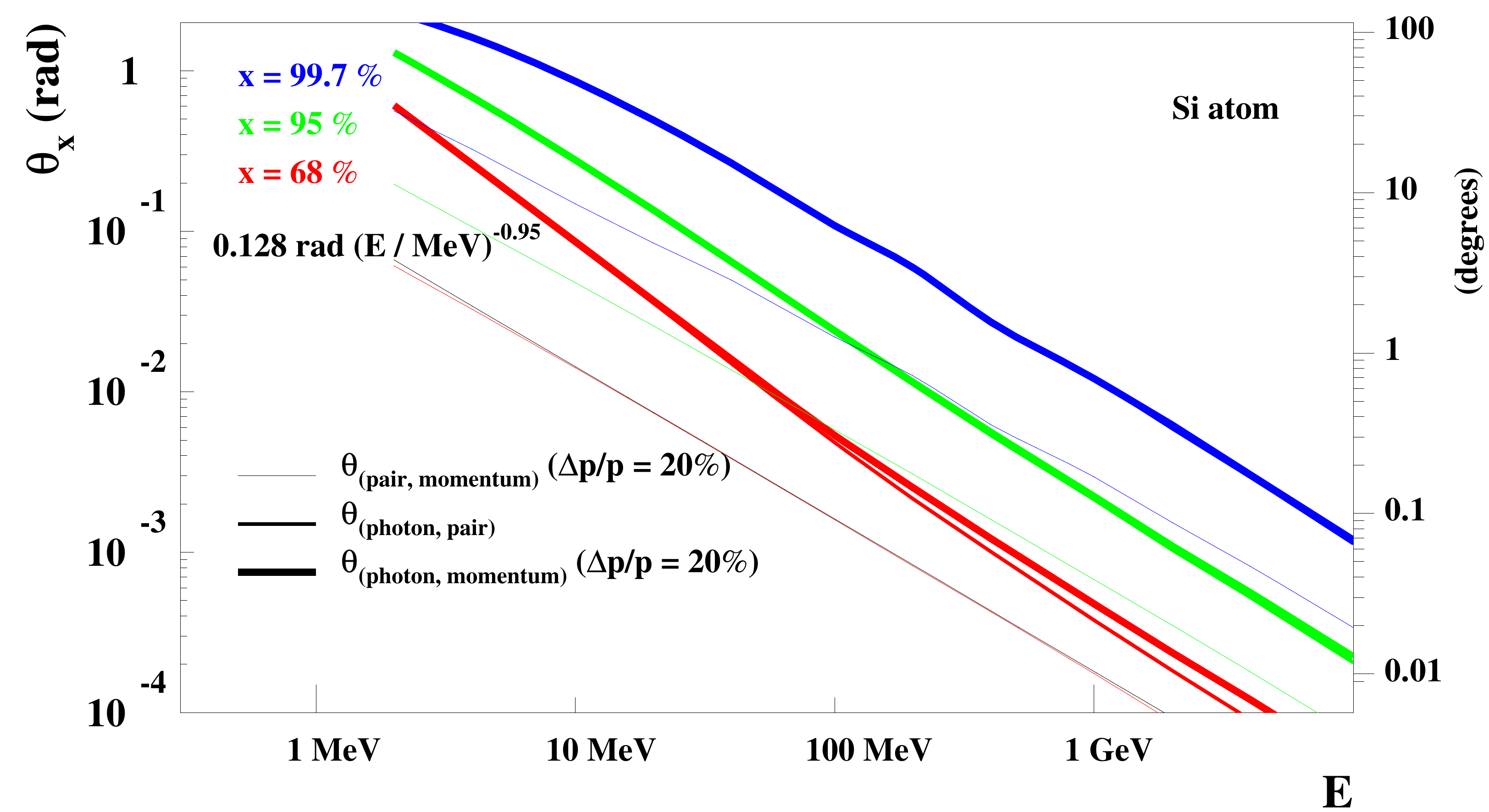}
\end{center}
\caption{\label{fig:mom} \sl 
Contribution to the single-photon angular resolution assuming a 
 $\sigma_p / p = 20\,\%$ single-track relative momentum resolution.
The thin black line is a good representation of the variation for $x = \Red{68}\,\%$.
}
\end{figure}

The track momenta can also be measured from the information collected
by the tracker/converter itself; as the RMS angle deflection undergone
from multiple scattering by a charged particle in its way through a
detector is inversely proportional to its momentum, the momentum can
be inferred from an analysis of the deflections 
\cite{Moliere,Kodama:2002dk}.
This method was used in emulsion-based $\gamma$ telescopes with a
typical relative momentum resolution of 10 to 20\,\%
\cite{Takahashi:2015jza}.

Assuming an RMS relative momentum resolution of $\sigma_p / p$ of
20\,\%, the contribution to the photon angular resolution is obviously
found to be better than when assuming equipartition.
%
%
 Figure \ref{fig:mom} shows
\begin{itemize}
\item the contribution of a 20\,\% relative momentum resolution 
 ($\theta_{(\paire,\momentum)}$, thin curve), and 
\item the already mentioned contribution of the missing recoil momentum
 ($\theta_{(\photon,\paire)}$, medium thickness)

 compared and combined, into
\item the full angular resolution, assuming a 20\,\% relative momentum
  resolution ($\theta_{(\photon,\momentum)}$, thick curve).
\end{itemize}
The 68\,\% containment value, that can be parametrized by
$0.64 \, \radian \left( {\sigma_p}/{p} \right) \left( {E}/{\mega\electronvolt} \right)^{-0.95}$,
is found to be  negligible when compared to the recoil
contribution, on the whole photon energy range.

\subsection{Interim wrap-up, (1)+(2)}

\begin{figure} [th]
\begin{center}
\includegraphics[width=0.82\linewidth]{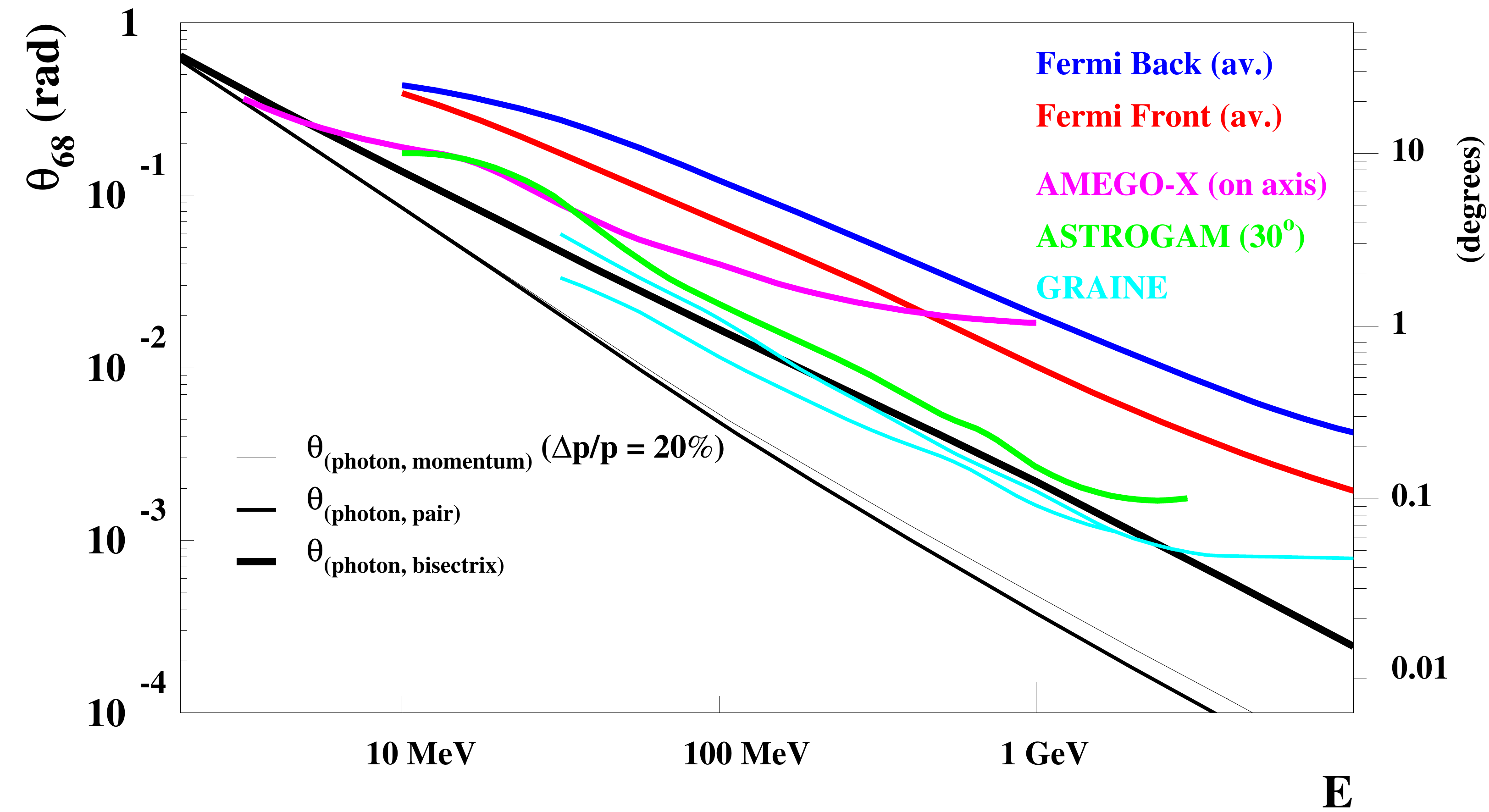}
\end{center}
\caption{\label{fig:wrap} \sl 
Telescope single-photon angular resolution as a function of $E$ for the
{\sl Fermi} LAT \cite{Fermi:LAT:Performance},
 GRAINE, \cite{Takahashi:2018nkq},
 AMEGO-X, \cite{Caputo:2022xpx} and
 ASTROGAM, \cite{e-ASTROGAM:2016bph}, compared to
 the full (including the recoil) resolution assuming a $\sigma_p / p = 20\,\%$ single-track relative momentum resolution, ($\theta_{(\photon,\momentum)}$, thick curve),
 the contribution of the missing recoil momentum ($\theta_{(\photon,\paire)}$, medium thickness) and
 the full resolution assuming equipartition ($\theta_{(\photon,\bisectrix)}$, thick curve).
}
\end{figure}

At this point, let's attempt an interim wrap-up comparing the
predictions for the two contributions examined up to now, (1)+(2), to
the angular resolutions of the present telescopes and future projects
(Fig. \ref{fig:wrap}).

\begin{itemize}
\item
The angular resolutions for all telescopes are found to be larger than the predictions.
\item
For the telescopes for which the photon direction was reconstructed
using the measured value of the track momenta
(GRAINE, eq. (3.1) of \cite{Ozaki:PhD};
ASTROGAM, eq. (1) of \cite{Aboudan:2022ver};
AMEGO-X, H. Fleischhack, private communication),
it is interesting to note that, instead, their resolutions would rather be compatible with the prediction for equipartition.
\end{itemize}

\clearpage

\subsection{Optimal track momentum measurement}

An optimal measurement of the track momentum can be obtained from a
bayesian analysis of the innovations of a Kalman-filter-based track
fit applied to the track \cite{Frosini:2017ftq}.
Let's consider a converter/tracker
consisting of a series of $N$ layers,
with a spacing $\ell$,
and scatterer thickness  $\Delta$ with radiation length $X_0$,
with a track position precision $\sigma$ for each layer;
(for an homogeneous detector, $\Delta = \ell$).

\begin{figure} [th]
\begin{center}
\includegraphics[width=0.2\linewidth]{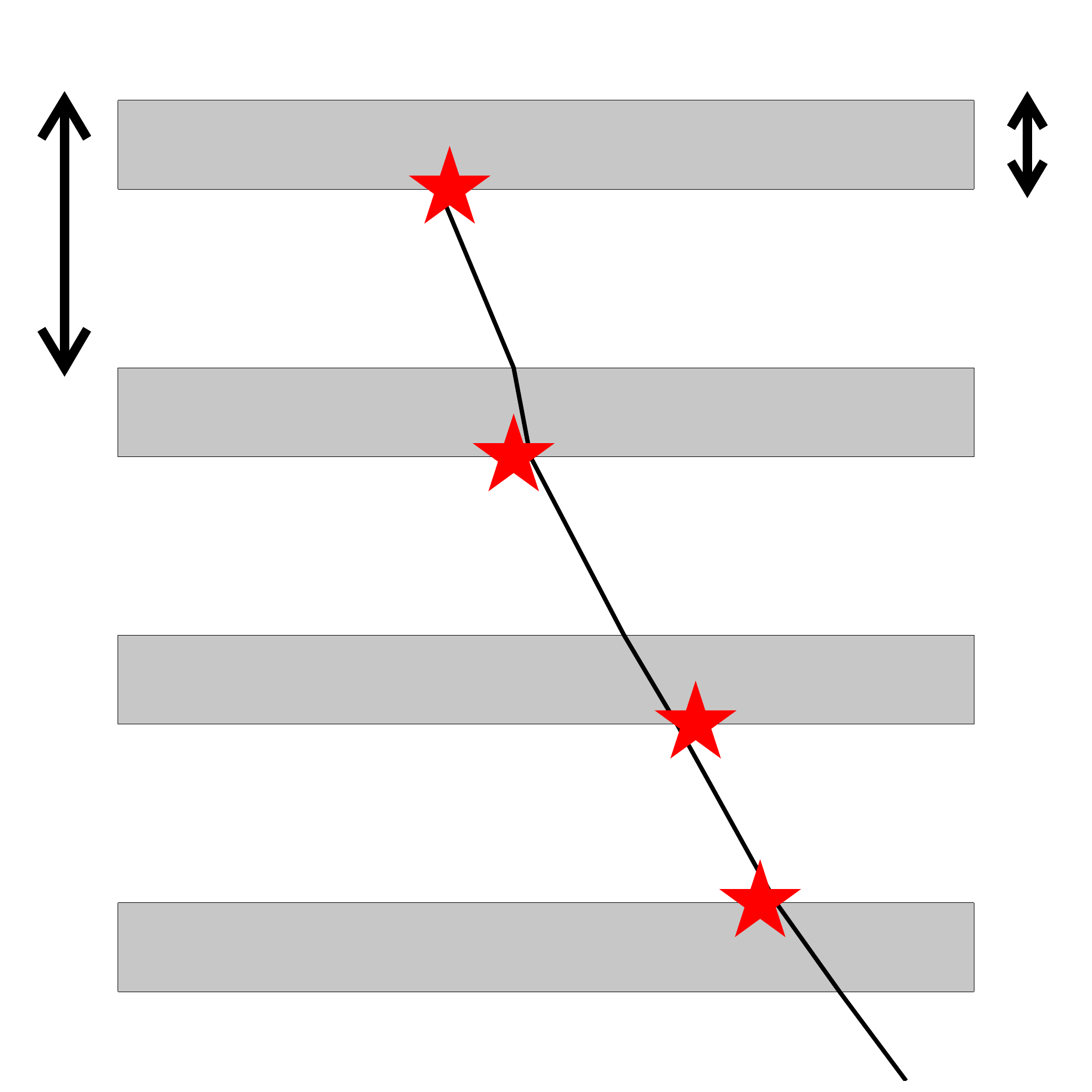}
\put(-195,77){layer spacing, $\ell$}
\put(10,85){$\Delta$, scatterer thickness}
\put(-91,24){vacuum}
\end{center}
\caption{\label{fig:tracker:converter:schema} \sl 
 Schema of a converter/tracker.
}
\end{figure}

\begin{figure} [th]
\begin{center}
 \includegraphics[width=0.49\linewidth]{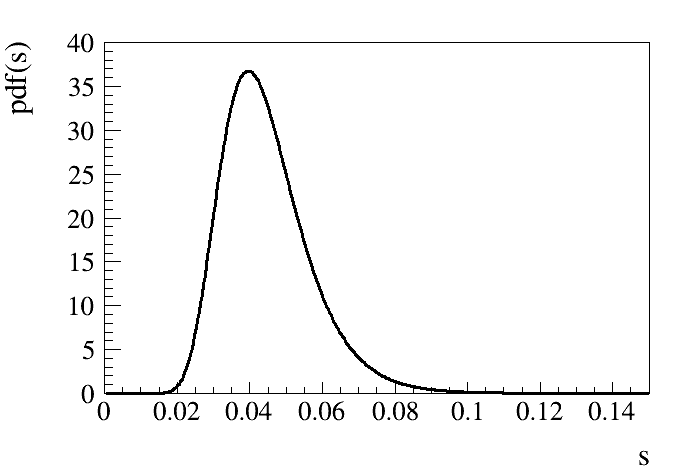}
\includegraphics[width=0.49\linewidth]{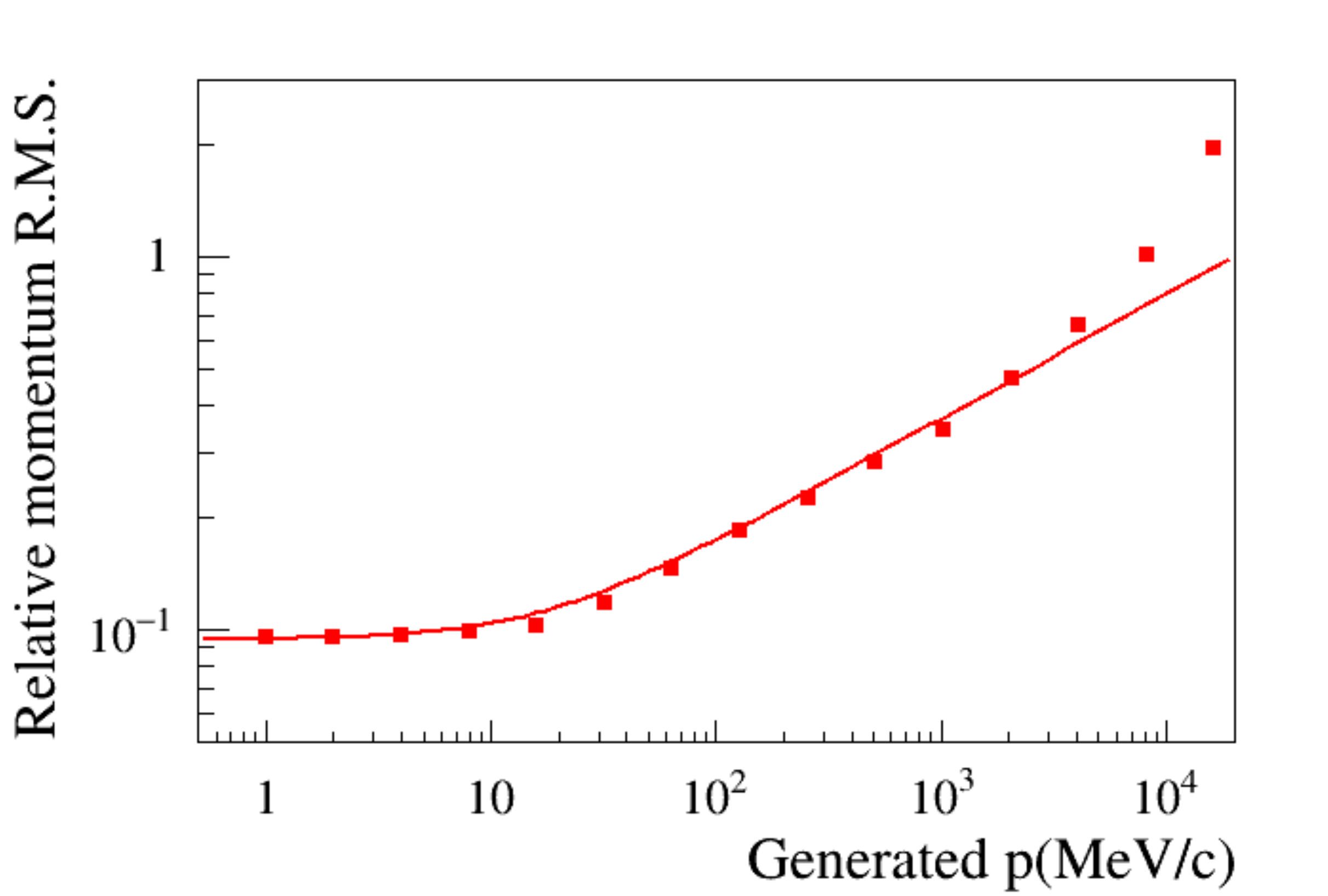}
\end{center}
\caption{\label{fig:opt:mom} \sl 
Internal (to the tracker/converter) optimal track-momentum measurement
from the analysis of the multiple deflections induced by multiple
scattering, for a DSSD tracker/converter (ASTROGAM)
 \cite{Frosini:2017ftq}.
Left, Kalman filter innovation probability density function as a
function of $s$, the average multiple-scattering angle variance per
unit track length, for a $50\, \mega\electronvolt / c$ track.
Right, relative momentum resolution as a function of (true) track momentum;
the curve shows the parametrization of eq. (\ref{eq:sigma:sur:p:parametrique}). 
}
\end{figure}

The value of the track momentum is obtained \cite{Frosini:2017ftq}
from the value of $s$, the average multiple-scattering angle variance
per unit track length,
\begin{equation}
s \equiv \left(\gfrac{p_0}{p}\right)^2 \gfrac{\Delta}{\ell X_0},
\end{equation}
that maximizes the Kalman filter innovation probability density function (Fig. \ref{fig:opt:mom}, left).
$p_0 = 13.6 \, \mega\electronvolt / c$ is the multiple scattering constant \cite{ParticleDataGroup:2022pth}.

A good representation of the relative precision of the measurement
(curve in Fig. \ref{fig:opt:mom}, right) is given by
(eq. (58) of \cite{Frosini:2017ftq})
\begin{equation}\label{eq:sigma:sur:p:parametrique}
 \gfrac{\sigma_p}{p} \approx \gfrac{1}{\sqrt{2 N}}
 \sqrt[4]{
1 + 256
 \left( \gfrac{p}{p_0} \right)^{4/3}
 \left( \gfrac{\sigma^2 X_0}{N \Delta ~ \ell^2} \right)^{2/3}
} ,
\end{equation}

A quantity of interest is, $p_{1/2}$, the momentum above which $\sigma_p/p$ is larger than
1/2, which means that the measurement becomes meaningless
 \cite{Frosini:2017ftq}:
\begin{equation}\label{eq:p1/2}
 p_{1/2} = p_0
 \gfrac{N^2}{32^{3/2}}
 \left( \gfrac{\ell} {\sigma}\right)
 \left( \gfrac{ \Delta} {X_0}\right)^{1/2}.
\end{equation}
For ASTROGAM, the method is usable on our whole energy range, with 
$p_{1/2} =2.5 \,\giga\electronvolt/c$ (Fig. \ref{fig:opt:mom}).

\begin{figure} [th]
\begin{center}
\includegraphics[width=0.55\linewidth]{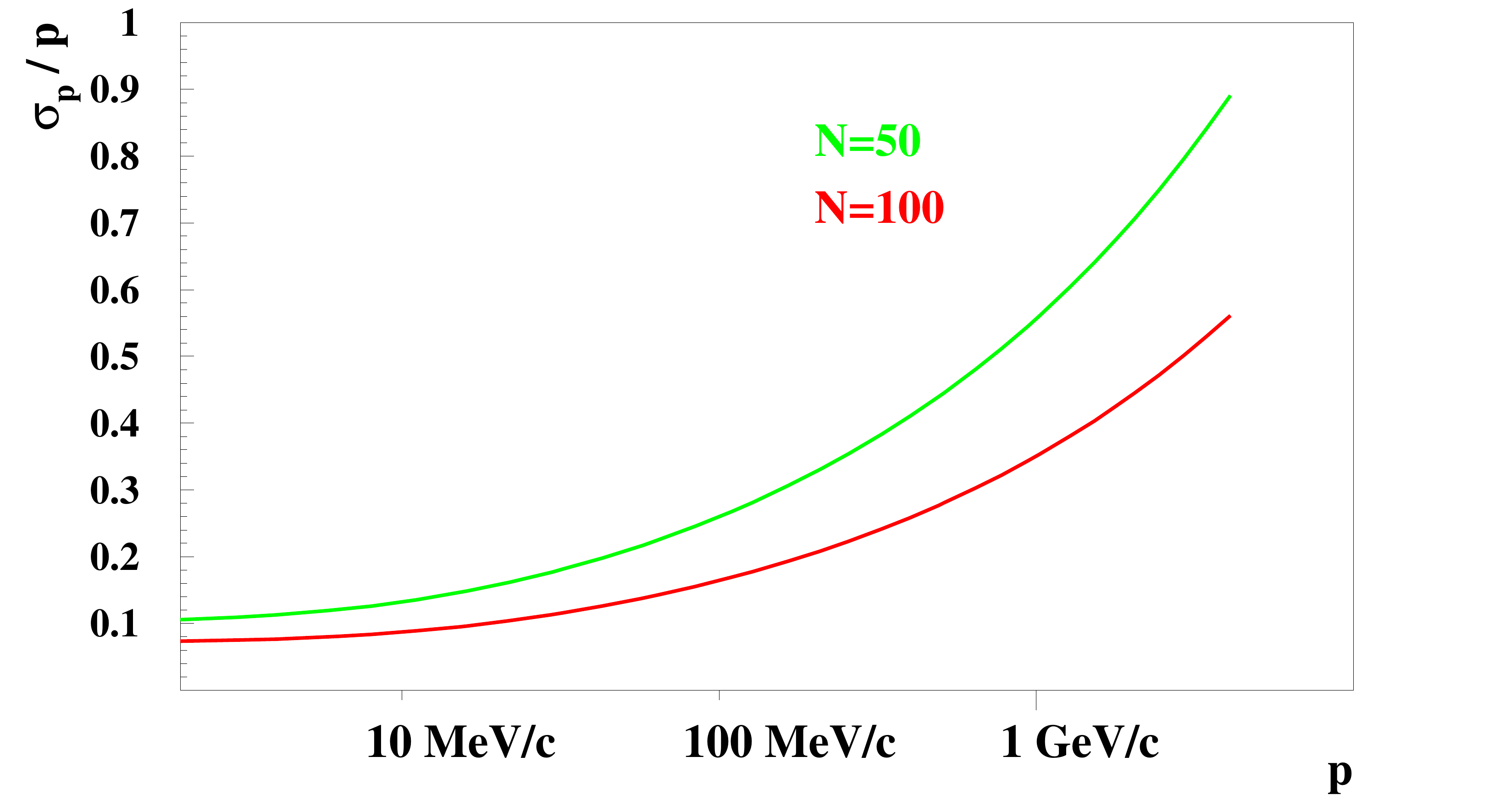}
\put(-230,124){\Green{\footnotesize \textbf{Conversion in middle}}}
\put(-230,111){\Red{\footnotesize \textbf{Conversion at top}}}
\end{center}
\caption{\label{fig:opt:mom:GRAINE} \sl 
Internal optimal track momentum measurement from the analysis of the
multiple deflections induced by multiple scattering, applied to the
emulsion-based tracker/converter of the GRAINE project, as computed
with eq. (\ref{eq:sigma:sur:p:parametrique}).}
\end{figure}

The method should be usable for the GRAINE tracker/converter too, on
most of the MeV energy range (Fig. \ref{fig:opt:mom:GRAINE}).

\clearpage

\section{Single-Track Angular Resolution}

I now examine the last (but not least, alas) contribution to the
single-photon angular resolution, that is, the single-track angular
resolution.
Tracking, the determination of a parametrization of the trajectory of
a track in a tracker, and in particular of its direction
at the vertex (the beginning of the track) is a challenging issue, as the
space resolution of each tracker layer and the multiple scattering
(MS) contribute together, and because MS deflections induce correlations in the
measured positions in layers downstream to the position where they
took place.
An optimal way out is the use of a Kalman filter applied to the track
measurements \cite{Fruhwirth:1987fm}, a recursive process that combines
\begin{itemize}
\item
an optimal estimate of the ``state'' of the track at layer $i$ (and its variance matrix),
\item
 and the measurement at layer $i+1$ (and its variance matrix),
\item
into an optimal estimate  of the ``state'' at layer $i+1$ (and its variance matrix).
\end{itemize}
In a scheme of successive
(scatter),
(measurement), 
(drift in empty space) steps,
we consider first a track originating from a vertex located at the
very bottom of a layer (Fig. \ref{fig:tracker:converter:schema}),
that is, (measurement) then (drift), and then only (scattering).
An analytical expression of the precision of such a tracking has been
obtained in \cite{Frosini:2017ftq}
(see the validation plot in Fig. 8 of \cite{Frosini:2017ftq} but the
published expression in \cite{Frosini:2017ftq} has a misprint).
We have
(eq. (1) of \cite{Bernard:2019znc})
\begin{equation} 
 \sigma_{\theta} =
 \gfrac{\sigma}{\ell}
 \sqrt{
 \frac
 {2 \, x^3 \, \left(\sqrt{4 j - x^2} + \sqrt{- 4 j -x^2} \right)}
 { \left(\sqrt{4 j - x^2} + j x \right) \left(\sqrt{- 4 j - x^2} - j x \right)}
 }
 , \label{eq:resolution:kalman}
\end{equation}
where $x$ is the distance between layers, $\ell$, normalized to the
detector scattering length $\lambda$
(defined in eq. (17) of \cite{Frosini:2017ftq}, see also \cite{Innes:1992ge}),
$x\equiv \ell/\lambda = \sqrt{({\ell}/{\sigma}) ({p_0}/{p}) \sqrt{{\Delta}/{X_0}}}$,
$\sigma$ is the precision of the measurement of the track position in a layer, 
and $j$ is the imaginary unit.
Contrary to appearances, $\sigma_{\theta}$ turns out to be a real
quantity, which is good.
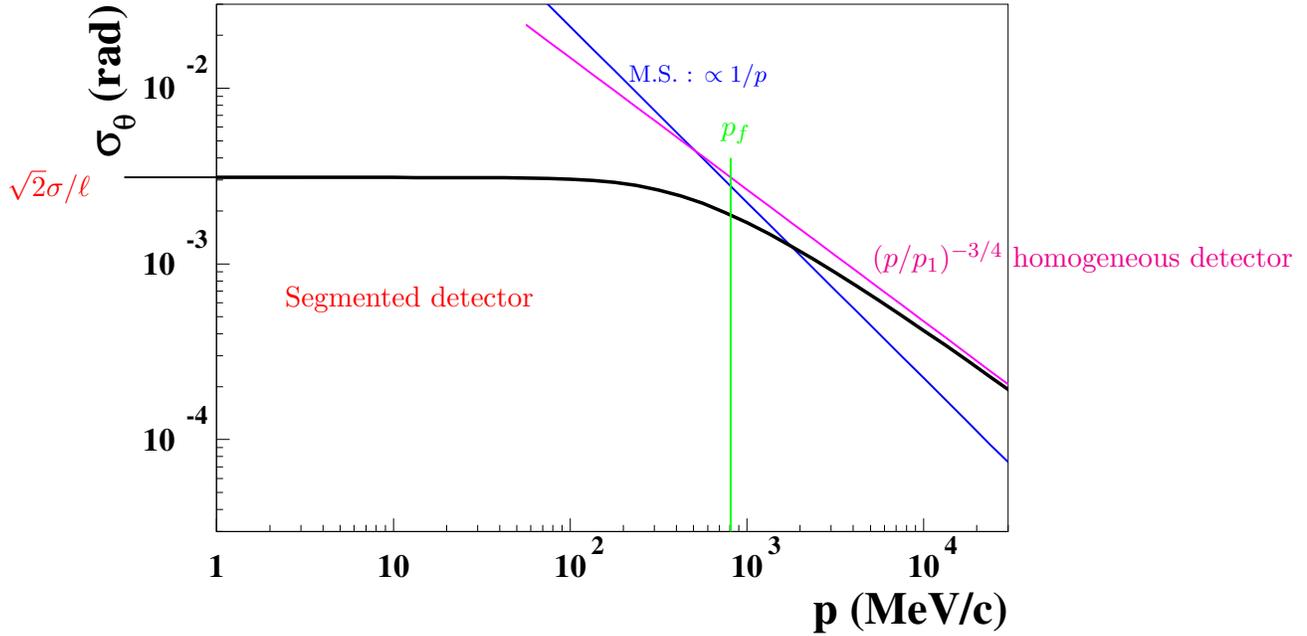
\begin{figure} [th]
\hspace{1cm}
 \includegraphics[width=0.7142857\linewidth]{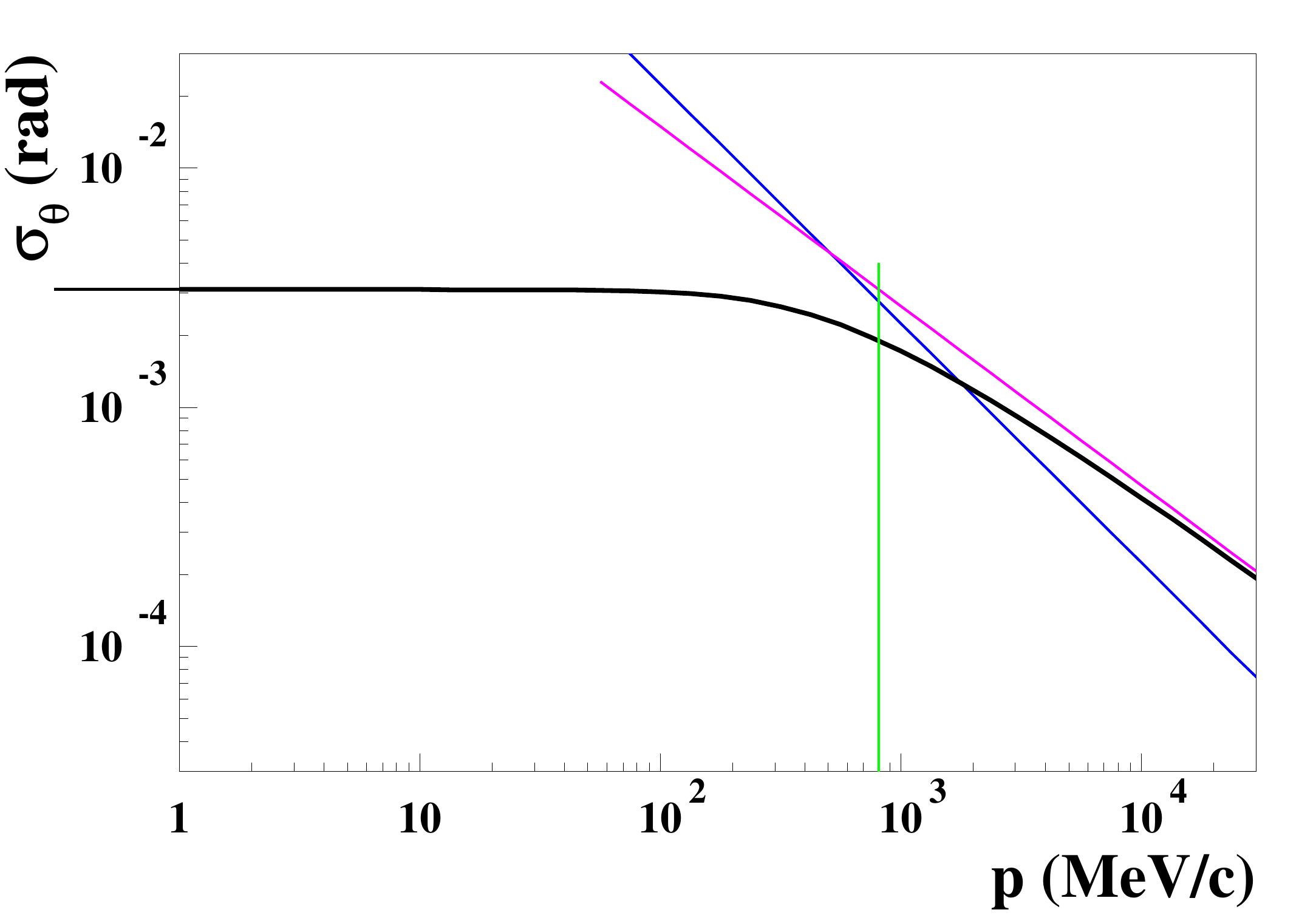}
 \put(-160,210){\Blue{\footnotesize M.S. : $\propto 1/p$}}
 \put(-395,167){\Red{$\sqrt{2} \sigma/\ell$}}
 \put(-125,190){\Green{$p_f$}}
 \put(-68,140){\Magenta{$({p}/{p_1})^{-3/4}$ homogeneous detector} }
 \put(-290,125){\Red{Segmented detector} }
 \caption{\label{fig:tracking} \sl
 Single track angular resolution for a vertex located at the very bottom of a layer, for the front part of the {\sl Fermi} LAT, as a function of track momentum.
 The black curve is from eq. (\ref{eq:resolution:kalman}), see also \cite{Bernard:2019znc}.
 The \Red{low momentum (segmented detector)} (eq. (\ref{eq:resolution:low:E})) and the
 \Magenta{high-momentum (homogeneous detector)} (eqs. (\ref{eq:resolution:high:E}), (\ref{eq:p1}) ) asymptotes are shown,
 together with the \Green{frontier} (vertical line, eq. (\ref{eq:pf}) ) between them.
 In addition, the RMS deflection from multiple scattering through a full layer (\Blue{M.S.}) is drawn. }
\end{figure}
The variation of $\sigma_{\theta}$ with $p$ (Fig.~\ref{fig:tracking})
shows two regimes\footnote{At
 much higher momenta, that is, outside the scope of this paper,
multiple scattering becomes negligible, and tracking boils down to the
fit of a straight line ($\vec B = 0$) or of an helix ($\vec B \ne
0$), with a single-track angular resolution at vertex of 
$ \sigma_{\theta t} \approx ({2 \sigma}/{L})\sqrt{3/(N+5)} $
and
$ \sigma_{\theta t} \approx ({8 \sigma}/{L})\sqrt{3/(N+5)} $,
respectively \cite{Regler:2008zza}.}:
\begin{itemize}
\item at low momentum (large $x$), the coarsely-segmented detector
 approximation can be used, with the obvious
\begin{equation} 
\sigma_{\theta} \approx \sqrt{2} \sigma/\ell
 , \label{eq:resolution:low:E}
\end{equation}
that is, 
an optimal measurement can be obtained simply from the position
measurements in the two first wafers, no Kalman filter is needed.
 
 \item at high momentum (small $x$), the power of the Kalman filter
 kicks in, and the homogeneous detector approximation \cite{Bernard:2013jea} can be used,
\begin{equation} 
 \sigma_{\theta} \approx \left({p}/{p_1}\right)^{-3/4}
 , \label{eq:resolution:high:E}
\end{equation}
where $p_1$ is a momentum that characterizes the
tracking-with-multiple-scattering properties of the detector
(eq. (17) of \cite{Bernard:2013jea},
eq. (38) of \cite{Frosini:2017ftq}).
\begin{equation}
 p_1 = p_0 \left( \gfrac{2 \sigma}{\ell} \right)^{1/3} \left(\gfrac{\Delta}{X_{0}} \right)^{1/2} .
 \label{eq:p1}
\end{equation}
\end{itemize}

The frontier between the realms of the segmented detector and the homogeneous
detector can be defined by the crossing of the two asymptotes, $p_f$
(vertical \Green{line} in Fig. \ref{fig:tracking})
\begin{equation} \label{eq:pf}
p_f = p_0 \sqrt{\gfrac{\Delta }{X_0}} \gfrac{\ell}{\sigma 2^{1/3}}.
\end{equation}

The numerical values of $p_1$, $p_{1/2}$ and of $p_f$ for the detector
test cases considered in this work can be found in Table
\ref{tab:dets}.

When conversion takes place within the fat of a layer, instead of the
very bottom of it, the contribution of multiple scattering in the
conversion layer must be taken into account in addition
(blue \Blue{line} in Fig. \ref{fig:tracking}, for a full layer).

\subsection{From single-track angular resolution to single-photon angular resolution }
 
The energy dependence of the single-photon angular resolution induced
by a momentum dependent single-track angular resolution has been
studied in Fig. 1 of \cite{Bernard:2012uf}.
Here I assume that the momentum dependence for the tracks translates
directly to the same energy dependence for the photons.

\clearpage

\subsection{Telescope Test Cases}

At this point, it's time to gather details on the tracker/converter of
the telescope test cases considered in this work (Table \ref{tab:dets}).
\begin{itemize}
\item
The tracker of the LAT consists of pairs of single-sided silicon
strip detectors, located just below a high-$Z$ material (tungsten)
foil, with thinner foils in an upper ``front'' set and thicker ones in
a ``back'' set \cite{Fermi-LAT:2009ihh}.
\item
The tracker of the ASTROGAM project consists of 56 planes of
double-sided silicon strip detectors
\cite{e-ASTROGAM:2016bph}.
\item
The ``historical'' AMEGO project has been similar to ASTROGAM, but in
its most recent version, AMEGO-X \cite{Caputo:2022xpx}, strips have
been changed to pixels, so as to decrease the capacitance of
individual detector elements, so that the low electronic noise enables
an improved sensitivity for very-low-energy Compton events.
This at the cost of an enlarged pitch of $1\milli\meter$.
\item
The tracker/converter of the GRAINE project consists of 100 films,
with a total thickness of 3\,cm, each film consisting of two emulsions
on both sides of a transparent base
\cite{Takahashi:2015jza,Ozaki:PhD}.
High-density emulsions are used with an average count density of
50\,counts\,/\,100\,microns.
Events can be reconstructed in two ways:
\begin{itemize}
\item
either films are scanned by an automatic machine, in which case the
track direction is computed from the position measured in the upper
and of the lower emulsion layers of the first
film\footnote{Multi-film measurements improve the angular resolution
 at GeV energies, i.e., outside the scope of this work, see Fig. 3 of
\cite{Takahashi:2018nkq}.}.
The relative positioning precision upon scanning is of about
$0.42\,\micro\meter$.
 The angular resolution is shown as the cyan \Cyan{upper curve} in Fig. \ref{fig:wrap}.
\item
or, the film is scanned by an operator with a microscope and the
position of each grain is determined with a precision of
$0.060 \,\micro\meter$;
The angular resolution is shown as the cyan \Cyan{lower curve} in Fig. \ref{fig:wrap}.
The small value of $p_1 = 45\, \kilo \electronvolt /c$ makes GRAINE well
suited for high-performance $\gamma$-ray polarimetry (see Fig. 20 of
\cite{Bernard:2013jea}) as demonstrated by the characterization of a
prototype on a polarized $\gamma$-ray beam \cite{Ozaki:2016gvw}.
\end{itemize}
\end{itemize}

\begin{table}
 \small
 \footnotesize
 \caption{Parameters of the text case telescopes considered in this work.
 \label{tab:dets}}
 \hspace{-1.5cm}
\begin{tabular}{l|c|c|c|c|c|c|l|l}
 \hline
 & \multicolumn{2}{c|}{{\sl Fermi} LAT \cite{Fermi-LAT:2009ihh}} & ASTROGAM \cite{e-ASTROGAM:2016bph} & AMEGO \cite{Caputo:2022xpx} & \multicolumn{2}{c|}{GRAINE \cite{Takahashi:2015jza,Ozaki:PhD}} & & \\
 \hline
 & \multicolumn{2}{c|}{ $2 \times$ SSSD + W foils} & DSSD & pixels & \multicolumn{2}{c|}{emulsions} & & \\
 & Front & Back & & & Automatic & Grain-by-grain & & \\
 $N$ & 12 & 4 & 56 & 60 & 100 & 35 (b) & & number of layers \\
 $\ell$ & 3 & 3 & 1 & 1 & & 0.0004 (b) & $\centi\meter$ & spacing between layers \\
 $\Delta$ & 95 & 720 & 500 & 500 & & 4 (b) & $\micro\meter$ & scatterer thickness \\
 & W & W & Si & Si & film & film & & scatterer nature \\
 $X_0$ & 0.35 & 0.35 & 9.37 & 9.37 & 2.84 & 2.84 & $\centi\meter$ & scatterer radiation length \\
 p & 228 & 228 & 240 & 1000 & & & $\micro\meter$ & pitch \\
 $\sigma$ & 66 (a) & 66 (a) & 70 (a) & 290 (a) & & 0.060 & $\micro\meter$ & single layer space resolution \\
 $p_1$ & 0.37 & 1.0 & 0.24 & 0.38 & & 0.045 & $\mega\electronvolt /c$ & detector ``scattering'' momentum \\
 & & & & & & & & eq. (\ref{eq:p1}) \\
 $p_{1/2}$ & 812. & 248. & 2460. & 684. & & 25.7 & $\mega\electronvolt /c$ & momentum for which $\sigma_p/p = 1/2$\\
 & & & & & & & & eq. (\ref{eq:p1/2}) \\
 $p_f$ & 810. & 2230. & 113. & 27. & & 3.0 & $\mega\electronvolt /c$ & frontier between the low and high \\ 
 & & & & & & & & energy asymptotes of the track \\
 & & & & & & & & angular resolution, eq. (\ref{eq:pf}) \\ 
 \hline
\end{tabular}

 (a) $\sigma = \pitch / \sqrt{12}$ 

(b) Poisson distributed, from 
the average count density of 50~counts~/~100~microns.
\end{table}

\begin{figure} [th]
\includegraphics[width=0.82\linewidth]{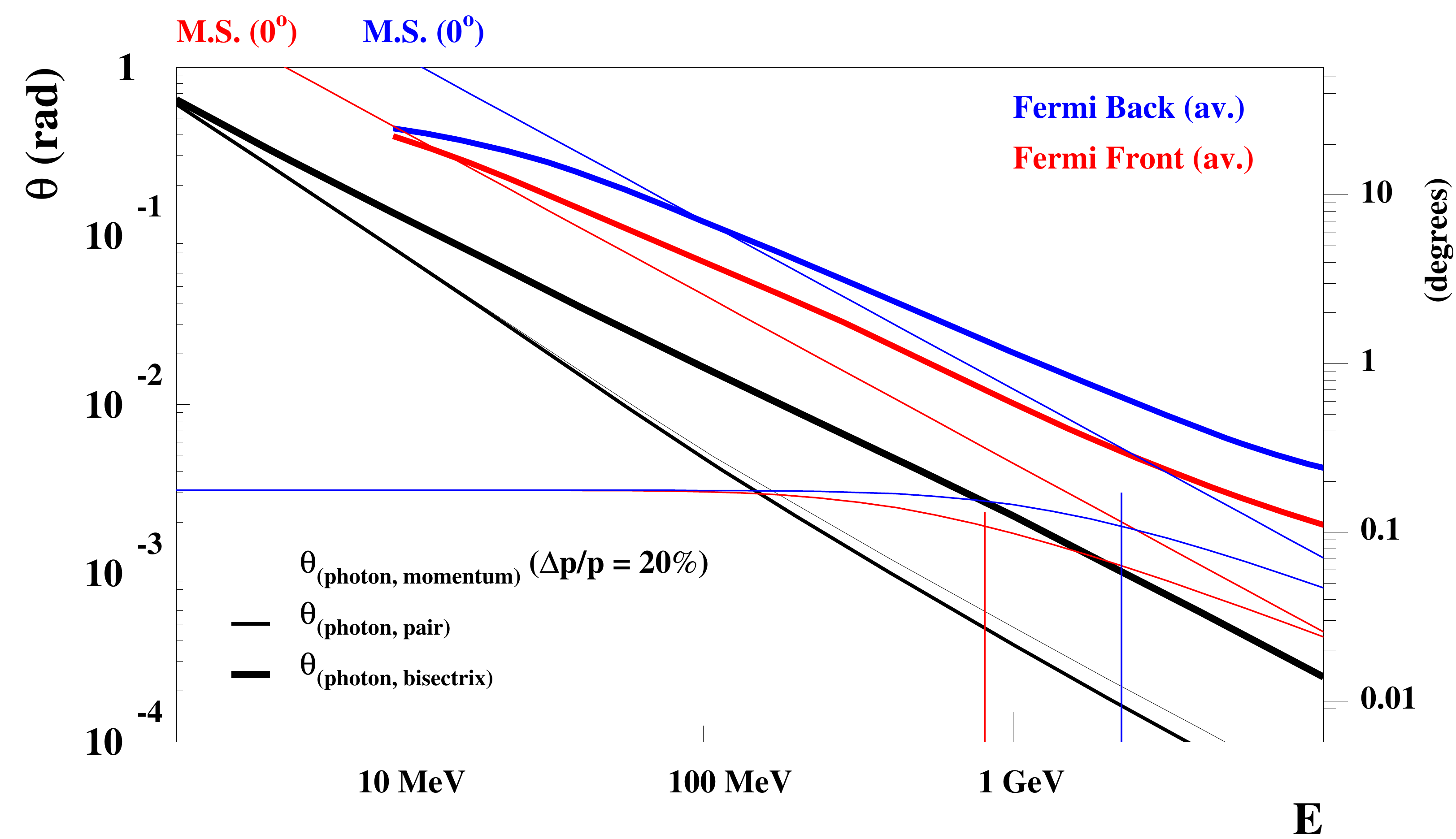}
\put(-28,63){\textbf{\footnotesize Kalman RMS (\Red{Front}, \Blue{Back})}}
\put(-336,239){\textbf{\footnotesize RMS ($p=E/2$, full conversion layer)}}
\put(-350,154){\textbf{\footnotesize 68\,\%}}
\put(-154,204){\textbf{\footnotesize 68\,\%}}
 \caption{\label{fig:tracking:fermi} \sl
The angular resolution of the {\sl Fermi} LAT (\Red{\bf front},
\Blue{\bf back}), averaged on the telescope solid angle acceptance
\cite{Fermi:LAT:Performance}, compared to the contributions studied in this work.
The ``bisectrix'' curve is shown even though the reconstruction of the LAT
data does not use that method.
The parametrization of the angular resolution of a Kalman filter on
track that would traverse the full (either \Red{\bf front} or
\Blue{\bf back}), computed from eq. (\ref{eq:resolution:kalman}), is
also drawn, together with the values of $p_f$ (vertical lines). }
\end{figure}

\begin{figure} [th]
\includegraphics[width=0.82\linewidth]{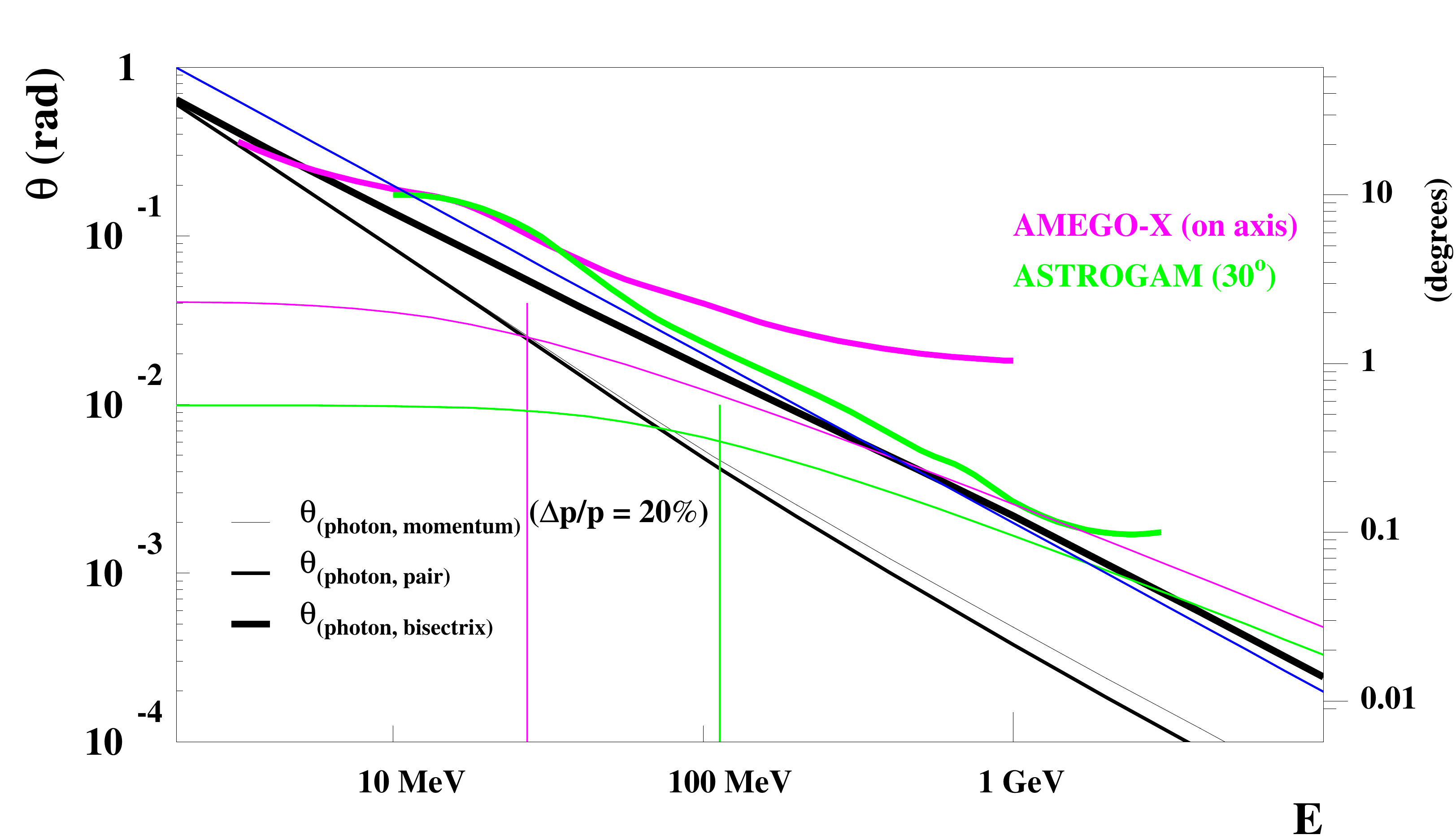}
\put(-28,63){\textbf{\footnotesize Kalman RMS (\Green{ASTROGAM})}}
\put(38,49){\textbf{\footnotesize (\Magenta{AMEGO-X})}}
\put(-336,210){\textbf{\Blue{\footnotesize MS RMS ($p=E/2$, full conversion layer)}}}
\put(-350,99){\textbf{\footnotesize 68\,\%}}
\put(-126,192){\textbf{\footnotesize 68\,\%}}
\caption{\label{fig:tracking:AA} \sl
The angular resolution of the
``pure silicon'' projects,
\Magenta{\bf AMEGO-X} and
\Green{\bf ASTROGAM}
 compared to the contributions studied in this work.
The ``bisectrix'' curve is shown even though the reconstruction of the LAT
data does not use that method.
The parametrization of the angular resolution of a Kalman filter on
track that would traverse the full detector,
 computed from eq. (\ref{eq:resolution:kalman}), is
also drawn, together with the values of $p_f$ (vertical lines). 
 }
\end{figure}

\begin{figure} [th]
\includegraphics[width=0.82\linewidth]{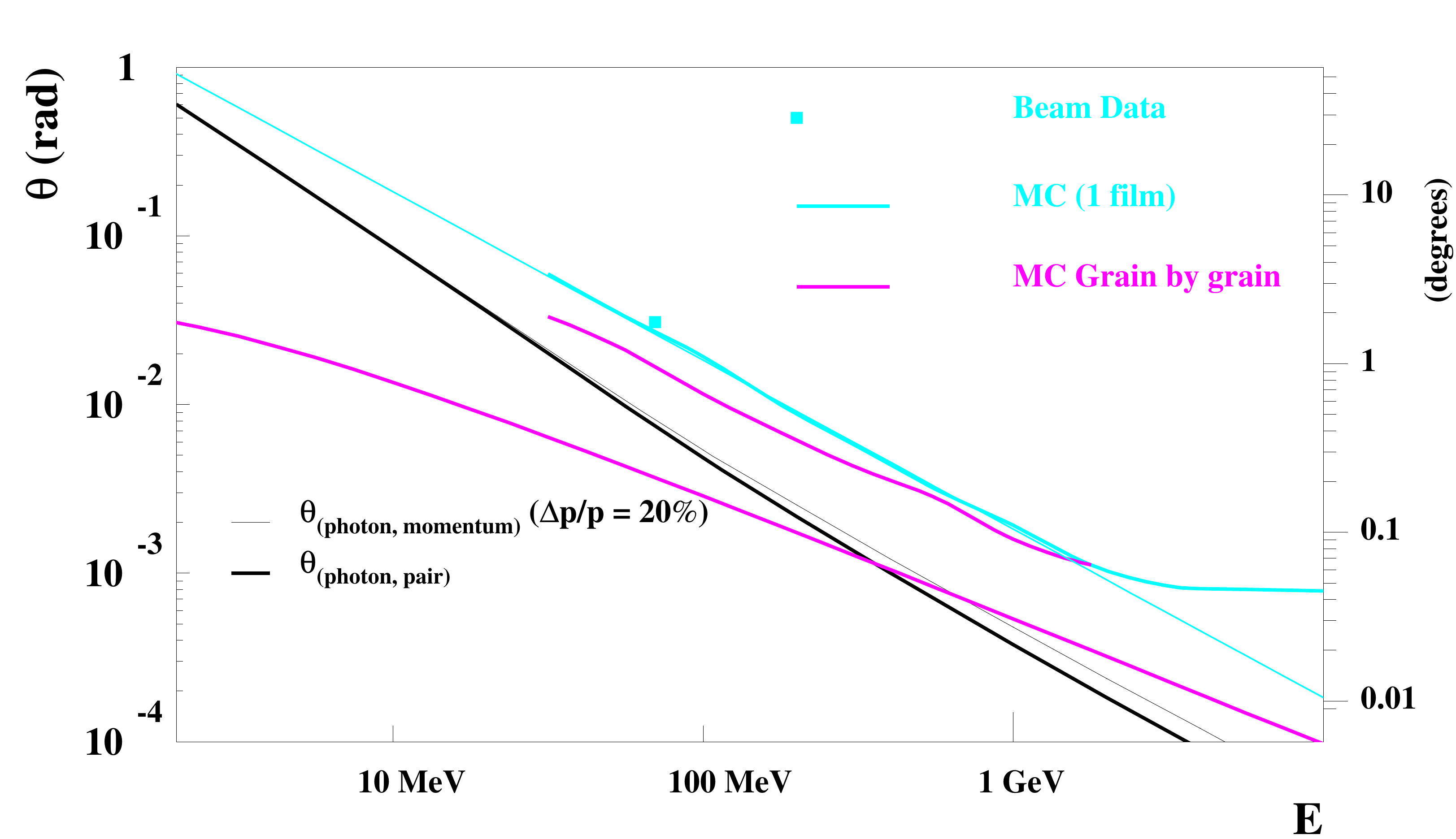}
\put(-28,203){\footnotesize $1.76 \pm 0.12 ^\circ$ @ $70.8\,\mega\electronvolt$ \cite{Ozaki:PhD}}
\put(-28,13){\textbf{\footnotesize \Magenta{Kalman RMS (Grain à grain})}}
\put(-336,210){\textbf{\footnotesize \Cyan{MS 1/3 film RMS}}}
\put(-350,99){\textbf{\footnotesize 68\,\%}}
\caption{\label{fig:tracking:GRAINE} \sl
The angular resolution of the GRAINE project, either with the
\Cyan{\bf automatic scanning} or with the \Magenta{\bf grain-by-grain} method,
compared to the contributions studied in this work.
The ``bisectrix'' curve is shown even though the reconstruction of the LAT
data does not use that method.
The parametrization of the angular resolution of a Kalman filter on
track that would traverse the full detector,
 computed from eq. (\ref{eq:resolution:kalman}), is
also drawn.
 }
\end{figure}

\section{Discussion}

\paragraph{{\sl Fermi} LAT } ~

The angular resolutions for the front and for the back part of the
{\sl Fermi} LAT (Fig. \ref{fig:tracking:fermi}) are vaguely compatible
with the multiple scattering inside the conversion layer (drawn for an
incident photon at $0^\circ$ and for a full layer), all other contributions 
considered in this work being completely negligible.
In particular the use of a Kalman filter would be useless on our
energy range.

It should be noted, though, that 
the slope (in log-log coordinate) of the data, close to $-0.8$, is
quite different from that of multiple scattering (that is, $-1$).
This might be due to the fact that the critical energy, $E_c$ of the
LAT, that is, the energy for which the limit pair opening angle that
can be measured from the two first layers, $\theta_c = \pitch /\ell$
is equal to the most probable pair opening angle
\cite{Bernard:2022jrj}, is equal to $211 \,\mega\electronvolt$, so the
transition between
\begin{itemize}
 \item
 a low energy regime for which most events have, in the second layer,
 the two tracks hit separate strips, and
 \item
a high-energy regime in which most events have the two tracks hit the
same strip,
\end{itemize}
falls right into our energy range.
Further more; the present event reconstruction, PASS 8
\cite{Fermi-LAT:2013jgq}, is strongly based on the reconstruction of
one main track, in a scheme somewhat different of what considered
here.

\paragraph{``Pure-silicon''} ~

The projects ASTROGAM and AMEGO-X, 
(Fig. \ref{fig:tracking:AA}),
are also dominated by multiple scattering in the conversion wafer
with a bizarre shape in the tens of MeV maybe related to the values of the critical energies
(67\,MeV for ASTROGAM; 16\,MeV for AMEGO-X);
(The bisectrix curve is drawn for reference but neither project is
using the method, as far as I know.)

As the Kalman filter is equivalent to a two-point measurement at low
energies
($p_f = 113\,\mega\electronvolt/c$ for ASTROGAM,
 $p_f = 27\,\mega\electronvolt/c$ for AMEGO-X),
the ability to assign the momenta (excellently) measured in the rest
of the event to the correct strip or pixel cluster in the second layer
is also key to a correct momentum weighting, similarly to the discussion of
the limits of track matching for the LAT at the end of section 6 of
\cite{Bernard:2022jrj}.

\paragraph{GRAINE} ~

For GRAINE, the performance of the automatic method is well understood from multiple
scattering in the first film (\Cyan{cyan curve} in Fig.
\ref{fig:tracking:GRAINE}).

With the grain-by-grain method, the tracker is so precise that the
performance should reach the kinematical limit up to a couple of
hundreds of MeV, but their \Magenta{results} are a bit larger than that.
The use of a Kalman filter may bring some improvement.

\section{Conclusion}

I studied several contributions to the single-photon angular
resolution of pair telescopes in the MeV energy range, by numerical methods.

The main strength of the study lies in the use of a validated sampling of the
exact, that is 5D, Bethe-Heitler differential cross section, 
which is key to the understanding of the distribution of the
(direction and magnitude) of the target recoil momentum, and
of the pair opening angle, the properties of which impact heavily the
photon angular resolution.

The contribution of the resolution of the magnitude of the momenta of
the tracks, $\sigma_p/p$ is found to be proportional to $\sigma_p/p$,
and for a mild assumption of $\sigma_p/p= 0.2$, is found to be smaller
than the kinematical limit of the unmeasured target recoil.

Multiple scattering was found to be the major issue, as was expected.
The use of a Kalman filter, in the hope of obtaining a better
determination of the photon direction than with a simple two-point
measurement, is found to be useless at low energies for silicon trackers
(below $\approx p_f$).
A Bayesian analysis of the filtering innovations of a Kalman filter,
though, provide an optimal measurement of track momenta, from the
multiple deflections of the tracks in the tracker due to multiple
scattering, that is sensitive on most of the energy range (below
$p_{1/2}$).

The emulsion-based GRAINE project stands out,
especially when the data are analyzed with the grain-by-grain method.
The tiny value of the detector scattering momentum, 
 $p_1 = 45\, \kilo \electronvolt /c$, paves the way to a high-quality
$\gamma$-ray polarimetry.
If a Kalman filter could be applied to the grain-by-grain data, the angular
resolution could most likely reach the kinematical limit.

A number of simplifying assumptions limit the scope of this study, and
a caveat is in order; radiation (bremsstrahlung) and energy loss of
charged particles in the tracker ($\dd E / \dd x$) are not considered,
something that has an important impact at the lowest energies (the assumption
of a track being measured in the full $N$ planes should be reconsidered);
also the effect of multiple scattering increases while the particle is
loosing energy.
Most often, conversions at the top of the tracker is implied, with a
contribution of the maximum number of layers, the same for all events.
And as mentioned in the text already, no pattern recognition algorithm
is used, so the transverse discretization of the reading of the signal
in strip- or pixel-segmented trackers, with the associated issue of a
pair of tracks possibly contributing to the same single cluster in the
second layer, is out of the scope of the present work based on the
presence of two well-identified tracks.
Also various definitions of the resolution, RMS or 68\,\%
containment value, have been used equivalently.

Never the less, I hope that the work will be of some use, in
particular as a guide for designers for future $\gamma$-ray
telescopes.

\section{Acknowledgements}

My gratitude to the organizers of the 2023 International Conference of
Deep Space Sciences for their kind invitation and to
Simone Maldera ({\sl Fermi} LAT),
Regina Caputo and Henrike Fleischhack (AMEGO-X), and
Satoru Takahashi (GRAINE)
for their help during the preparation of my talk.

\tableofcontents

\clearpage

\begin{tabular}{llllll}
 $\Aeff$ & effective area
 \\
 $B(E)$ & background differential flux 
 \\
 $\Delta E$ & incident photon energy bin
 \\
 $\Delta$ & scatterer thickness 
 \\
 $E$ & incident photon energy 
 \\
 $E_c$ & critical photon energy 
 \\
 $k$ & incident photon momentum
 \\
 $\ell$ & distance between layers (i.e. track longitudinal sampling)
 \\
 $\lambda$ & detector scattering length 
 \\
 $m$ & electron mass
 \\
 $M$ & nucleus mass
 \\
 $n$ & number of standard deviations (in definition of the sensitivity)
 \\
 $N$ & number of layers of the tracker/converter
 \\
 $p$ & momentum 
 \\
 $q$ & ``recoil'' momentum transferred to the target
 \\
 $s$ & sensitivity
 \\
 $s$ & the average multiple-scattering angle variance per unit track length
 \\
 $\sigma$ & R.M.S.
 \\
 $T$ & duration
 \\
 $\theta$ & polar angle (respective to the direction of the incident photon)
 \\
 $\theta_{+-}$ & pair opening angle
 \\
 $x$ & containment fraction
 \\
 $x$ & distance between wafers normalized to the
detector scattering length, $x\equiv \ell/\lambda$
 \\
 $X_0$ & radiation length
\end{tabular}


\clearpage

\begin{thebibliography}{00}
 \small

\bibitem{Rando:2022zyj}
R.~Rando,
``The {\sl Fermi} Large Area Telescope,''
 invited chapter for Handbook of X-ray and Gamma-ray Astrophysics (Eds. C. Bambi and A. Santangelo, Springer Singapore, 
\href{https://arxiv.org/abs/2208.13635}
 {[arXiv:2208.13635 [astro-ph.IM]]}.

\bibitem{Fermi-LAT:2022byn}
S.~Abdollahi \textit{et al.} [Fermi-LAT],
``Incremental {\sl Fermi} Large Area Telescope Fourth Source Catalog,''
\href{https://inspirehep.net/literature/2020890}{Astrophys. J. Supp. \textbf{260} (2022) 53}

\bibitem{Fermi:LAT:Performance}
 ``{\sl Fermi} LAT Performance'',
 Pass 8 Release 3 Version 3, 
(\href{https://www.slac.stanford.edu/exp/glast/groups/canda/archive/pass8r3v3/lat_Performance.htm}{html})
 
\bibitem{integral}
C. Winkler \textit{et al.}, 
``The INTEGRAL mission'',
\href{https://inspirehep.net/literature/633500}
{A\&A 411 (1) L1-L6 (2003)}

\bibitem{e-ASTROGAM:2017pxr}
A.~De Angelis \textit{et al.} [e-ASTROGAM],
``Science with e-ASTROGAM: A space mission for MeV\textendash{}GeV gamma-ray astrophysics,''
\href{https://inspirehep.net/literature/1634531}
{JHEAp \textbf{19} (2018), 1-106}

\bibitem{Bernard:2013jea} 
 D.~Bernard,
 ``Polarimetry of cosmic gamma-ray sources above $e^+e^-$ pair creation threshold'',
 \href{https://inspirehep.net/literature/1242601}
 {Nucl.\ Instrum.\ Meth.\ A {\bf 729} (2013) 765}.

\bibitem{COMPTEL}
 \href{https://heasarc.gsfc.nasa.gov/docs/cgro/epo/posters/Greatest_Hits/COMPTEL_1-30_MeV.html}
{1 to 30 MeV All-Sky Map}, 
CGRO Science Support Center.

\bibitem{LAT-20-200}
 L. Marcotulli {\it et al.}, [The Fermi LAT Collaboration],
 ``Bridging the Gap - The first sensitive 20-200 MeV catalog'',
Tenth International {\sl Fermi} Symposium, Johannesburg South Africa,
9-15 Oct 2022
(\href{https://indico.cern.ch/event/1091305/contributions/5007665/}{html})

\bibitem{Bernard:2012uf}
 D.~Bernard,
 ``TPC in gamma-ray astronomy above pair-creation threshold,''
 \href{https://inspirehep.net/literature/1198968}
 {Nucl.\ Instrum.\ Meth.\ A {\bf 701} (2013) 225},
 Erratum: [Nucl.\ Instrum.\ Meth.\ A {\bf 713} (2013) 76].

\bibitem{Gros:2016zst}
 P.~Gros and D.~Bernard,
 ``$\gamma$-ray telescopes using conversions to electron-positron pairs: event generators, angular resolution and polarimetry,''
 \href{https://inspirehep.net/literature/1504915}
 {Astropart.\ Phys.\ {\bf 88} (2017) 60}.
 
\bibitem{Aboudan:2022ver}
A.~Aboudan \textit{et al.}, 
``A Multimission Method for the Reconstruction of Gamma-ray Events on Silicon Tracker Pair Telescopes,''
\href{https://inspirehep.net/literature/2064204}
{Astrophys. J. \textbf{928} (2022) no.2, 141}

\bibitem{Olsen1963}
 H. Olsen, 
``Opening Angles of Electron-Positron Pairs'',
\href{https://inspirehep.net/literature/46674}
{Phys. Rev. {\bf 131} (1963) 406}.


\bibitem{Bethe-Heitler}
H. Bethe and W. Heitler,
``On the Stopping of Fast Particles and on the Creation of Positive Electrons'',
\href{https://inspirehep.net/literature/9146}
{Proceedings of the Royal Society of London A, {\bf 146} (1934) 83}.

\bibitem{BerlinMadansky1950}
 T. H. Berlin and L. Madansky,
``On the Detection of gamma-Ray Polarization by Pair Production'', 
 \href{https://doi.org/10.1103/PhysRev.78.623}
 {Phys. Rev. {\bf 78} (1950) 623}.

\bibitem{May1951}
 M. M. May,
``On the Polarization of High Energy Bremsstrahlung and of High Energy Pairs'',
 \href{https://doi.org/10.1103/PhysRev.84.265}{Phys. Rev. {\bf 84} (1951) 265}.

\bibitem{Bernard:2018hwf}
 D.~Bernard,
 ``A 5D, polarised, Bethe-Heitler event generator for $\gamma \to e^+e^-$ conversion'',
 \href{https://inspirehep.net/literature/1657191}{Nucl.\ Instrum.\ Meth.\ A {\bf 899} (2018) 85}
 
\bibitem{Semeniouk:2019cwl}
I.~Semeniouk and D.~Bernard,
``C++ implementation of Bethe-Heitler, 5D, polarized, $\gamma{\to}e^+e^-$ pair conversion event generator,''
\href{https://inspirehep.net/literature/1741571}
{Nucl. Instrum. Meth. A \textbf{936} (2019) 290}

\bibitem{Geant4:2019cxv}
V.~Ivanchenko \textit{et al.} [Geant4],
``Progress of Geant4 electromagnetic physics developments and applications,''
\href{https://inspirehep.net/literature/1760905}
{EPJ Web Conf. \textbf{214} (2019) 02046}

\bibitem{Agostinelli:2002hh}
 S.~Agostinelli {\it et al.} [GEANT4 Collaboration],
 ``GEANT4: A Simulation toolkit,''
\href{https://inspirehep.net/literature/593382}
{Nucl.\ Instrum.\ Meth.\ A {\bf 506} (2003) 250}.

\bibitem{Allison:2016lfl}
 J.~Allison {\it et al.},
 ``Recent Developments in Geant4,''
\href{https://inspirehep.net/literature/1488031}
 {Nucl.\ Instrum.\ Meth.\ A {\bf 835} (2016) 186}.
 
\bibitem{Jost:1950zz} 
 R.~Jost \textit{et al.}, 
 ``Distribution of Recoil Nucleus in Pair Production by Photons,''
\href{https://inspirehep.net/literature/47140}
 {Phys.\ Rev.\ {\bf 80} 189 (1950) 189}.

\bibitem{Borsellino1953}
 A. Borsellino, 
``Momentum Transfer and Angle of Divergence of Pairs Produced by Photons'',
\href{http://prola.aps.org/abstract/PR/v89/i5/p1023_1}
{Phys. Rev. {\bf 89} (1953) 1023}.

\bibitem{Caputo:2022xpx}
R.~Caputo \textit{et al.},
``All-sky Medium Energy Gamma-ray Observatory eXplorer mission concept,''
\href{https://inspirehep.net/literature/2134383}
{J. Astron. Telesc. Instrum. Syst. \textbf{8} (2022) no.4, 044003}

\bibitem{e-ASTROGAM:2016bph}
A.~De Angelis \textit{et al.} [e-ASTROGAM],
``The e-ASTROGAM mission,''
\href{https://inspirehep.net/literature/1495966}
{Exper. Astron. \textbf{44} (2017) no.1, 25}
 
\bibitem{Beischer:2020rts}
B.~Beischer,
``Measurement of High Energy Gamma Rays from 200 MeV to 1 TeV with the Alpha Magnetic Spectrometer on the International Space Station,''
 CERN-THESIS-2020-065 
[arXiv:2007.08392 [astro-ph.HE]].
 \href{https://inspirehep.net/literature/1807286}
 {Ph D (2020)}

\bibitem{Moliere}
 G. Molière,
 ``Theorie der Streuung schneller geladener Teilchen. III. Die Vielfachstreuung von Bahnspuren unter Berüksichtigung der statistischen Kopplung'',
 \href{https://doi.org/10.1515/zna-1955-0301}{Zeitschrift Naturforschung A {\bf 10} (1955) 177}.

\bibitem{Kodama:2002dk}
K.~Kodama \textit{et al.}
``Detection and analysis of tau neutrino interactions in DONUT emulsion target,''
 \href{https://inspirehep.net/literature/603038}
{Nucl. Instrum. Meth. A \textbf{493} (2002) 45}

\bibitem{Takahashi:2015jza} 
 S.~Takahashi {\it et al.},
 ``GRAINE project: The first balloon-borne, emulsion gamma-ray telescope experiment,''
\href{https://inspirehep.net/literature/1362123}
{PTEP {\bf 2015} (2015) 043H01}.


\bibitem{Ozaki:PhD}
 K. Ozaki,
``Development of an emulsion telescope system for gamma-ray source observations and realization of the GRAINE 2015 balloon-borne experiment in Australia'', 
 \href{https://da.lib.kobe-u.ac.jp/da/kernel/D1006569/?lang=1}
{D1006569},
{Ph. D.} 2016, Kobe University



\bibitem{Takahashi:2018nkq}
S.~Takahashi \textit{et al.} [GRAINE],
``GRAINE project, prospects for scientific balloon-borne experiments,''
\href{https://inspirehep.net/literature/1699801}
{Adv. Space Res. \textbf{62} (2018), 2945}


\bibitem{Frosini:2017ftq}
 M.~Frosini and D.~Bernard,
 ``Charged particle tracking without magnetic field: optimal measurement of track momentum by a Bayesian analysis of the multiple measurements of deflections due to multiple scattering,''
 \href{https://inspirehep.net/literature/1605726}
 {Nucl.\ Instrum.\ Meth.\ A {\bf 867} (2017) 182}.

 
\bibitem{ParticleDataGroup:2022pth}
R.~L.~Workman \textit{et al.} [Particle Data Group],
``Review of Particle Physics,''
 \href{https://inspirehep.net/literature/2106994}
{PTEP \textbf{2022} (2022) 083C01}

\bibitem{Fruhwirth:1987fm}
 R.~Frühwirth,
 ``Application of Kalman filtering to track and vertex fitting,''
 \href{https://inspirehep.net/literature/259509}
{Nucl.\ Instrum.\ Meth.\ A {\bf 262} (1987) 444}. 

\bibitem{Bernard:2019znc}
D.~Bernard,
``Performance of the MeV gamma-ray telescopes and polarimeters of the future. $\gamma \to e^+ e^-$ in silicon-detector active targets,''
 \href{https://inspirehep.net/literature/1721134}
{Mem. Soc. Ast. It. \textbf{90} (2019) 149}

\bibitem{Innes:1992ge}
W.~R.~Innes,
``Some formulas for estimating tracking errors,''
 \href{https://inspirehep.net/literature/335574}
{Nucl. Instrum. Meth. A \textbf{329} (1993) 238}

\bibitem{Regler:2008zza}
M.~Regler and R.~Fruhwirth,
``Generalization of the Gluckstern formulas. I: Higher orders, alternatives and exact results,''
 \href{https://inspirehep.net/literature/790377}
{Nucl. Instrum. Meth. A \textbf{589} (2008) 109}




\bibitem{Fermi-LAT:2009ihh}
W.~B.~Atwood \textit{et al.} [Fermi-LAT],
``The Large Area Telescope on the Fermi Gamma-ray Space Telescope Mission,''
 \href{https://inspirehep.net/literature/812754}
{Astrophys. J. \textbf{697} (2009) 1071}


\bibitem{Ozaki:2016gvw}
K.~Ozaki \textit{et al.},
``Demonstration of polarization sensitivity of emulsion-based pair conversion telescope for cosmic gamma-ray polarimetry,''
\href{https://inspirehep.net/literature/1455851}
{Nucl. Instrum. Meth. A \textbf{833} (2016) 165}




\bibitem{Bernard:2022jrj}
D.~Bernard,
``MeV-GeV Polarimetry with $\gamma \to e^+e^-$: Asserting the Performance of Silicon Strip Detectors-Based Telescopes,''
\href{https://inspirehep.net/literature/2146414}
{Nucl. Instrum. Meth. A \textbf{1042} (2022) 167462}

\bibitem{Fermi-LAT:2013jgq}
W.~Atwood \textit{et al.} [Fermi-LAT],
``Pass 8: Toward the Full Realization of the Fermi-LAT Scientific Potential,''
2012 Fermi Symposium proceedings -
\href{https://inspirehep.net/literature/1223837}
{eConf C121028},
[arXiv:1303.3514 [astro-ph.IM]].
\end{thebibliography}
\end{document}